\documentclass[trackchanges,twocolumn]{aastex701}
	\usepackage{amsmath}

\begin{document}
			
			\title{Probing the accretion geometry of the transient accreting millisecond pulsar SAX J1808.4-3658: a possible transition to the propeller regime}

			\author[orcid=0000-0002-4004-2323]{Mahasweta Bhattacharya}
			\affiliation{Visva-Bharati University, Santiniketan, West Bengal, 731235, India}
			\email{03333372303@visva-bharati.ac.in} 
			
			\correspondingauthor{Mahasweta Bhattacharya$^{1*}$}
			\email{E-mail: 03333372303@visva-bharati.ac.in$^{1*}$\newline 
				adityas.mondal@visva-bharati.ac.in$^{1}$\newline
				gulabd@iucaa.in$^{2}$}

			\author[orcid=0000-0002-2565-1219]{Aditya S. Mondal} 
			\affiliation{Visva-Bharati University, Santiniketan, West Bengal, 731235, India}
			\email{adityas.mondal@visva-bharati.ac.in}
			
			\author[orcid=0000-0003-1589-2075]{Gulab C. Dewangan}
			\affiliation{Inter-University Centre for  Astronomy \& Astrophysics (IUCAA), Pune, 411007, India}
			\email{gulabd@iucaa.in}
			
			
			
			
			
			
			\begin{abstract}
				
				We analyze three \textit{NuSTAR} observations and two \textit{NICER} observations of the transient accreting millisecond pulsar SAX J1808.4-3658 in the hard spectral state during its most recent outbursts in 2022 and 2025. The spectral analysis shows that the persistent continuum is well described by an absorbed thermal Comptonization model, with electron temperatures of $\sim$48--92 keV and $\sim$48--110 keV for the 2022 and 2025 observations, respectively. Prominent iron emission line $\sim$5--8 keV and a Compton hump $\sim$15--30 keV are detected in all \textit{NuSTAR} observations, indicating reflection of hard X-ray photons from the accretion disc. Reflection modeling with \texttt{relxillCP} indicates a moderate inclination of $\sim$30$^{\circ}-$60$^{\circ}$. During the 2025 September observation, the inferred inner radius of 25$^{+17}_{-16}$$R_{\rm g}$ (at 1$\sigma$ uncertainty level) corresponds to the magnetospheric radius exceeding the co-rotation radius, consistent with the possibility that the source was approaching the propeller regime. Although the disc is truncated at the larger radius, accreted material is still reaching the surface of the neutron star, which is confirmed through the detection of a type-I X-ray burst during this \textit{NuSTAR} observation. The spectral analysis of the burst suggests helium burning at a low ignition depth.
				
			\end{abstract}
			
		
		
		\section{Introduction} 
		
		Neutron star low mass X-ray binaries (NS LMXBs) are binary systems comprising a neutron star (NS) that accretes gas from a donor companion of mass $M \lesssim 1 M_{\odot}$ (less massive than the compact primary; \citealt{1997LNP...487....1T, Bahramian_2022, DiSalvo2023aoc}). Accreting millisecond X-ray pulsars (AMXPs) are NS LMXBs where the accreted matter spins up the NS with high frequency ($\nu\gtrsim100$ Hz) and the accreted gas gets channeled out of the accretion disc by weak magnetic fields (B$\sim$10$^{8-9}$ G) onto the magnetic poles of the neutron star, giving rise to the X-ray pulsations at the spin frequency \citep{2021ASSL..461..143P}. Transient AMXPs exhibit bright, sudden outbursts driven by the inflow of accreted matter from the companion star, and these systems are thought to form a link between the LMXBs and isolated radio MSPs \citep[e.g., ][]{2018cosp...42E2854R}. During outbursts, variations in the spin frequency are governed by the net angular momentum transferred via the interaction between the accreting plasma and the magnetic field \citep{2017ApJ...835....4B, 2016ApJ...818...49M, Sanna2020}. The NS spin-up involves a mechanism called the \textit{recycling scenario}, in which accretion from the companion transfers angular momentum to the NS. A key prerequisite for reaching millisecond periods is a low magnetic field, facilitated by the dissipation of the strong magnetic field ($\sim 10^{12}$~G) to a weaker state ($\sim 10^8$~G) \citep{1982Natur.300..728A, 1982CSci...51.1096R}.\\
		
		In LMXBs, the matter from the companion star is transferred via Roche-lobe overflow and accreted onto the NS through the accretion disc \citep{1987huba.conf...35B}. Classically, the inner edge of the accretion disc is defined by the magnetospheric radius ($r_{\rm m}$), where the pressure of the accretion disc balances the magnetic pressure of the NS \citep{ghosh1979accretiona}. In the disc-magnetosphere interaction model, the star's magnetic field is governed by a magnetic torque produced by field lines threading the disc over a range of radii around the magnetospheric radius \citep{ghosh1979accretion}. The inner radius where the disc begins to couple strongly to the stellar magnetic field scales as $r_{\rm m} \propto \mu^{4/7} \dot{M}^{-2/7}$, where $\mu=BR^3$ is the magnetic moment for a NS with magnetic field $B$ and radius $R$, and $\dot{M}$ is the mass accretion rate. The co-rotation radius ($r_{\rm co}$) is defined as the location where the Keplerian angular velocity of the accretion disc equals the spin angular velocity of the NS \citep{owocki2009stellar}. If the magnetosphere resides within the co-rotation radius ($r_{\rm m} < r_{\rm co}$), the standard accretion mechanism proceeds \citep{hartmann1999comparisons}. In this state, the relatively strong magnetic field channels the accreting matter along the magnetic field lines towards the magnetic poles of the NS, producing coherent pulsed emissions characteristic of AMXPs \citep{2008AIPC.1054..173D}. Broad relativistic iron emission lines observed in AMXP spectra provide a valuable tool to constrain the magnetospheric radius and estimate the strength of the magnetic field in the NS \citep{Fabian1989, Cackett_2009}. However, recent studies suggest that the inner disc may truncate before reaching the classical magnetospheric radius \citep{vandenEijnden2016, marino2022outflows}.\\
		As the mass accretion rate ($\dot{M}$) decreases, the ram pressure of the inflowing matter of the accretion disc drops, allowing the magnetosphere to expand outward ($r_{\rm m}\propto\dot{M}^{-2/7}$) \citep{stella1984magnetic, 1986ApJ...305..235T}. Once the magnetospheric boundary expands beyond the co-rotation radius ($r_{\rm m} > r_{\rm co}$), the system enters the propeller regime \citep{1976ApJ...207..914A, 2016ApJ...817..100P}. At this interface, the inflowing plasma interacts with magnetic field lines rotating at a velocity higher than the Keplerian angular velocity of the accretion disc matter \citep{hawley1999transport}. The centrifugal force due to the spin angular velocity exceeds the gravitational pull of the NS \citep{morbidelli2018accretion}. A centrifugal barrier occurs at the magnetospheric boundary that prevents inflow of the accreted matter \citep{lyutikov2023centrifugal}. This centrifugal inhibition of accretion results in the ejection of matter and is called the propeller stage of accretion \citep{ustyugova2006propeller}. \\

		The source SAX J1808.4-3658 (hereafter referred to as SAX J1808) was discovered during the 1996 observation of \textit{BeppoSAX} using Wide Field Cameras \citep{zand1998discovery}. The 401 Hz pulsations detected from the \textit{RXTE} instrument established the source as an AMXP \citep{1998Natur.394..344W}. It is the first detected AMXP with a 2.01 hr orbital period \citep{chakrabarty1998two}. The iron emission line has been confirmed in previous observations of SAX J1808, which were used to assess the inner edge of the accretion disc \citep{Miller_2007, Cackett_2009}.\\
		SAX J1808 shows outbursts every 2-4 years; the outburst analysed in this work was reported in the Astronomer's Telegram \text{ATel}$\# 17323$ and  $\# 17324$ \citep{2019MNRAS.483..767D, 2020ApJ...905...87B, 2023ApJ...942L..40I, 2025ATel17323....1R, 2025ATel17324....1W}.\\
		
		The peak flux of SAX J1808 during the 1996 outburst is reported as $\sim$2.1, $\sim$4.0, and $\sim$3.3$\times$10$^{-9}$ erg cm$^{-2}$ s$^{-1}$ in the energy bands 2-10 keV, 2-28 keV, and 3-25 keV, respectively \citep{cornelisse2001first}. After the main outburst, the luminosity of the source decreases, and it is said to enter a reflare phase. \cite{campana2008swift} reported changes in the accretion flow during the 2005 reflare phase. The flux and magnetic field during the 2008 outburst were estimated as $(1.74\pm0.02)\times10^{-9}$ erg cm$^{-2}$ s$^{-1}$ and $B=(3.2\pm1.0)\times10^8$ G, assuming a distance to the source of $d$=$3.5\pm0.1$ kpc \citep{2009ApJ...702.1673H, Cackett_2009}. The inner disc radius and inclination of the accretion disc were estimated as $R_{\rm in}=(13.2\pm2.5) GM c^{-2}$ and $i=55^{\circ} {}^{+8^{\circ}}_{-4^{\circ}}$, respectively. The flux reached a minimum of $2.0\times10^{-13}$ erg cm$^{-2}$ s$^{-1}$ in the 0.5-10 keV band during the reflaring phase after the 2008 outburst \citep{Patruno_2009}. This behavior is consistent with the onset of accretion flow instabilities associated with the transition of SAX J1808 into the propeller regime. These variations are likely driven by changes in the accretion geometry. In the 2015 outburst, the 0.6-10 keV flux, inner disc radius, and inclination estimated were $\sim2\times10^{-9}$ erg cm$^{-2}$ s$^{-1}$, $\sim$8.2 $GM c^{-2}$, and $68^{\circ}\pm4^{\circ}$, respectively \citep{2019MNRAS.483..767D}. SAX J1808 has exhibited eleven $\sim$1-month-long outbursts; the tenth one initiated on 19 August 2022 \citep{2026ApJ...996...73B, 2026ApJ...999..133S}. During the reflaring stage following the 2022 outburst, the 0.5-10 keV flux was reported to be $4.24_{-0.10}^{+0.14}\times10^{-11}$ erg cm$^{-2}$ s$^{-1}$, which is one of the faintest reported luminosities for the source \citep{Ballocco_2025}. The inner radius was estimated to be close to the co-rotation radius. Their study suggested that the magnetospheric radius ($r_{\rm m}$) remained close to the co-rotation radius ($r_{\rm co}$), indicating the onset of the propeller regime. Despite this, the low X-ray luminosity of the source implied that a small fraction of the accreted material continued to leak onto the NS surface, indicating the accretion was not completely suppressed. Such low-level accretion is thought to proceed through a modified disc configuration such as a `trapped' or `dead' disc, rather than the standard Shakura-Sunyaev disc \citep{patruno2016reflares}.\\
		Moreover, type-I X-ray bursts have been observed from this source on many occasions \citep{zand1998discoveryxraytransientsax, cornelisse2001first, 2019MNRAS.483..767D, Bult_2020}. The burst flux for the bursts observed in 2002 outbursts was in the range $\sim(1.60-1.82)\times10^{-7}$ erg cm$^{-2}$ s$^{-1}$ \citep{galloway2006helium}. A precursor is manifested as a short spike before the rising phase of the main burst. The first evidence of a precursor in a normal hydrogen-helium powered thermonuclear burst was reported during the 2002 October 19 burst \citep{bhattacharyya2007unusual}.\\

		Motivated by \cite{Ballocco_2025}, we undertook a spectral analysis to investigate the geometric and physical properties of the inner disc during the 2025 outburst. In particular, a type-I burst was detected during the 2025 September observation, while the spectral properties suggested that the source may have been approaching the propeller regime. In this present work, we have analyzed three \textit{NuSTAR} observations of the accreting millisecond pulsar SAX J1808, taken during two different outbursts in 2022 and 2025. We have re-analyzed the 2022 \textit{NuSTAR} observation to ensure a consistent comparison among the three observations. In particular, all datasets were reduced using the same procedures and calibration files, and the same spectral models and fitting strategy were applied throughout. By re-analyzing the 2022 \textit{NuSTAR} observation within the same analysis framework as the 2025 \textit{NuSTAR} observations, we ensure that the observed changes in the continuum and reflection parameters are directly comparable. Moreover, we performed a joint spectral fit of simultaneous 2022 \textit{NuSTAR} and \textit{NICER} observations to provide more robust constraints on the fundamental spectral parameter. This paper is structured as follows. In Section \ref{sec:obs_data}, we state the observation details and describe the data reduction process. Throughout Section \ref{sec:lc}, we discuss the properties of the light curves corresponding to the observations. In Section \ref{sec:spec_analys}, we present spectral analysis techniques adapted for the related models corresponding to the persistent and burst emission to carry out the spectral analysis. Finally, in Section \ref{sec:disc}, we carry out the discussion based on the observations and inferences drawn from the spectral analysis.\\

		\begin{figure*}
			\centering
			\includegraphics[width=0.40\linewidth, height=5.2cm]{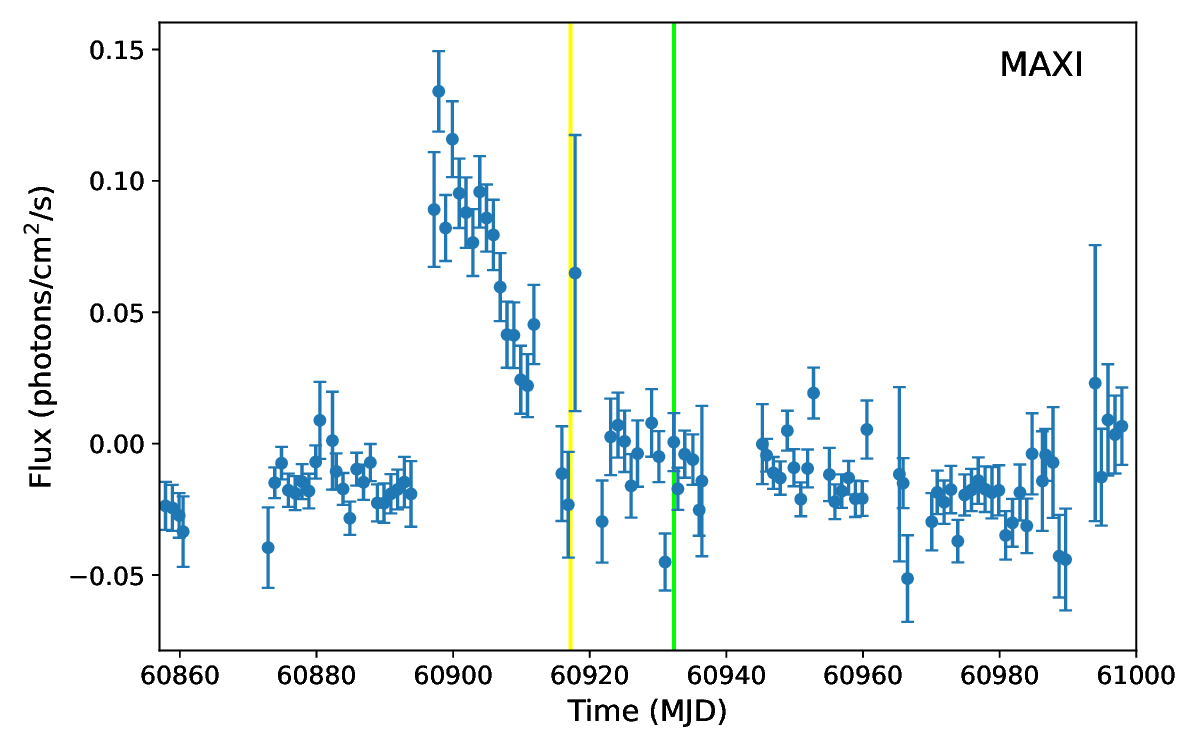}
			\includegraphics[width=0.40\linewidth]{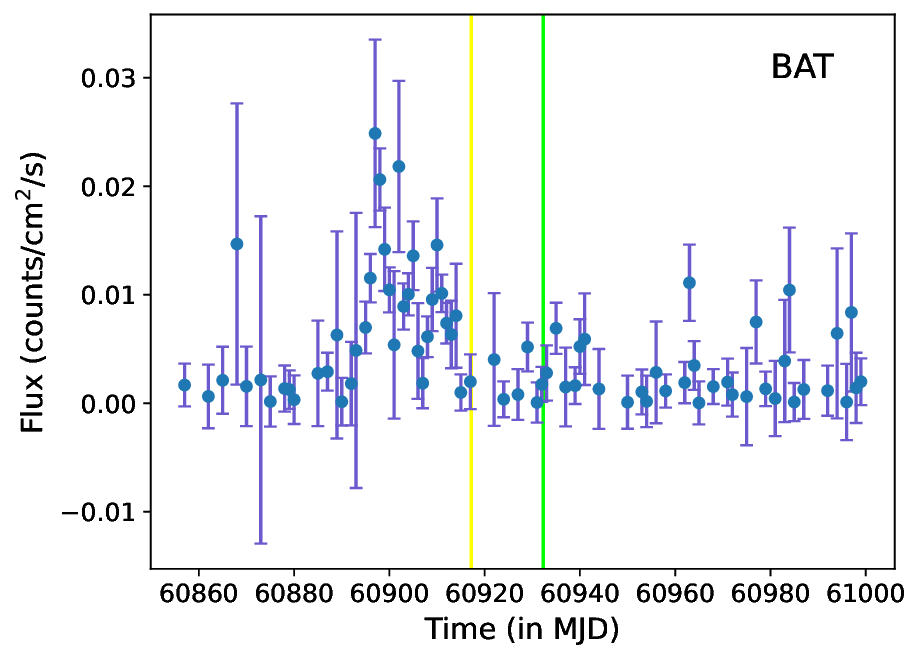}
			\caption{Light curve for the source SAX J1808 for 2025 \textit{NuSTAR} observations corresponding to observation IDs 91101333002 and 91101333004 that took place in the decay phase of the 2025 outbursts are shown separately as \textit{yellow} and \textit{lime} colors, respectively for the \textit{left panel:} \textit{MAXI} light curve in 2-6 keV energy range, \textit{right panel:} \textit{Swift}/BAT light curve in the 15-50 keV energy regime.}
			\label{fig:maxi_data}
		\end{figure*}
		
		\section{Observation and Data reduction}
		\label{sec:obs_data}
		\textit{Nuclear Spectroscopic Telescopic Array} (\textit{NuSTAR}), launched by NASA on 2012 June 13, is the first focusing hard X-ray telescope operating in the high energy ($3-79$ keV) range (\citealt{harrison2013nuclear}). \textit{Neutron Star Interior Composition Explorer} (\textit{NICER}) mission is dedicated towards the study of thermal and non-thermal emission from neutron star in the 0.2-12 keV (soft) X-ray band \citep{2016SPIE.9905E..1HG}.
		
		This work is based on three \textit{NuSTAR} observations of SAX J1808 obtained during the 2022 and 2025 outbursts. We also include two \textit{NICER} observations conducted before and after the 2022 \textit{NuSTAR} observation \citep{2023ApJ...942L..40I}. Details of all observations are listed in Table \ref{tab:obs_all}. Hereafter, we refer to the \textit{NuSTAR} and \textit{NICER} observation IDs as labeled in Table \ref{tab:obs_all}, as Obs$\sim$1-5, unless stated otherwise.\\
		
		\begin{table*}
			\centering
			\caption{Details of the observations IDs of the source SAX J1808 that are utilized in this work.}
			\begin{tabular}{l c l l c}
				\hline
				\parbox{2.3cm}{Observation ID \\ (Label)}	& Date (MJD) & Instrument & \parbox{1.2cm}{Exposure \\ (ks)} & \parbox{2 cm}{Net Count Rate \\ (counts s$^{-1}$)}\\
				\hline
				80701312002 (Obs 1) & 2022.08.22 (59813.72) & \textit{NuSTAR FPMA/B} & 107 & 16 \\
				91101333002 (Obs 2) & 2025.08.30 (60917.06) & \textit{NuSTAR FPMA/B} & 19.3 & 6 \\
				91101333004 (Obs 3) & 2025.09.14 (60932.19) & \textit{NuSTAR FPMA/B} & 21.2 & 4 \\
				5050260104  (Obs 4) & 2022.08.21 (59812.99) & \textit{NICER/XTI} & 9.4 & 272 \\
				5050260105  (Obs 5) & 2022.08.23 (59814.02) & \textit{NICER/XTI} & 7.2 & 275 \\
				\hline
			\end{tabular}
			\label{tab:obs_all}
		\end{table*}
		
		The \textit{NuSTAR} data were processed using the data analysis software \texttt{NuSTARDAS v0.4.12}, within \texttt{HEASOFT v6.36} employing the latest calibration database (\texttt{CALDB v20251215}). The task \texttt{nupipeline v0.4.12} was used to generate the calibrated and screened event files. For all the observations, circular extraction region of 100'' radius was used to study the persistent and burst spectrum. From the same chip, the background was selected as a region of radius 100'' far away from the source. HEASOFT provides FTOOLS to deduce the FITS files needed for analysing the observed data. FTOOL \texttt{nuproducts} was used to extract the spectra and light curves from FPMA and FPMB. For time-resolved spectroscopy, good time interval (GTI) files were created separately for the persistent emission and the observed Type-I X-ray burst. \textbf{We have set the minimum counts per bin (\texttt{grpmincounts}) during the \texttt{nuproducts} procedure to 30 and 25 for the persistent emission spectra and burst spectra, respectively \citep{Kaastra_2016}.} During burst analysis, data above 20 keV were excluded due to background dominance. The \textit{NICER} standard calibration, screening and filtering of the data is done using the \texttt{nicerl2 v1.41} pipeline \citep{remillard2022empirical}. The light curves were produced using \texttt{nicerl3-lc} task. The screened files obtained from \texttt{nicerl2} pipeline are further utilised to extract the required files for spectral analysis using \texttt{nicerl3-spect} task. Additionally, the background was modelled using the default \textsc{SCORPEON} background model implemented in \texttt{nicerl3-spect}.\\

		\section{Light Curve}
		\label{sec:lc}
		
		\begin{figure*}
			\centering
			\includegraphics[width=0.6\columnwidth]{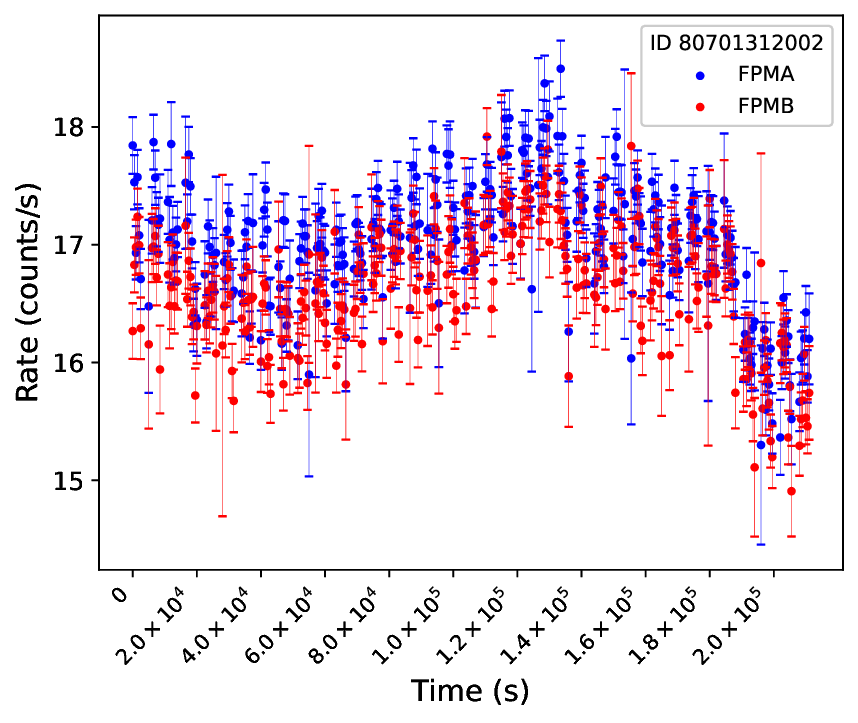}
			\includegraphics[width=0.6\columnwidth]{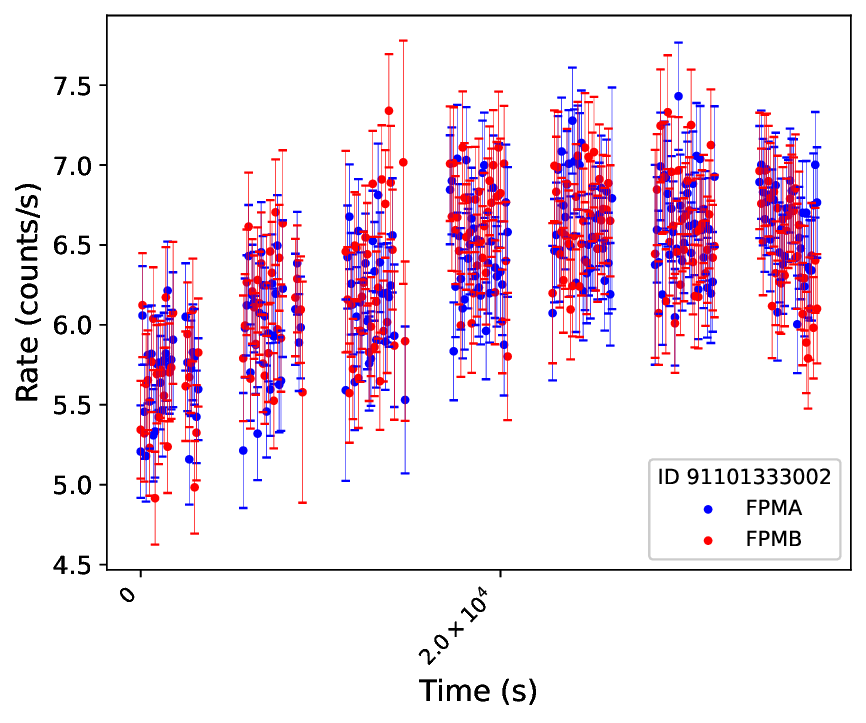}
			\includegraphics[width=0.6\columnwidth]{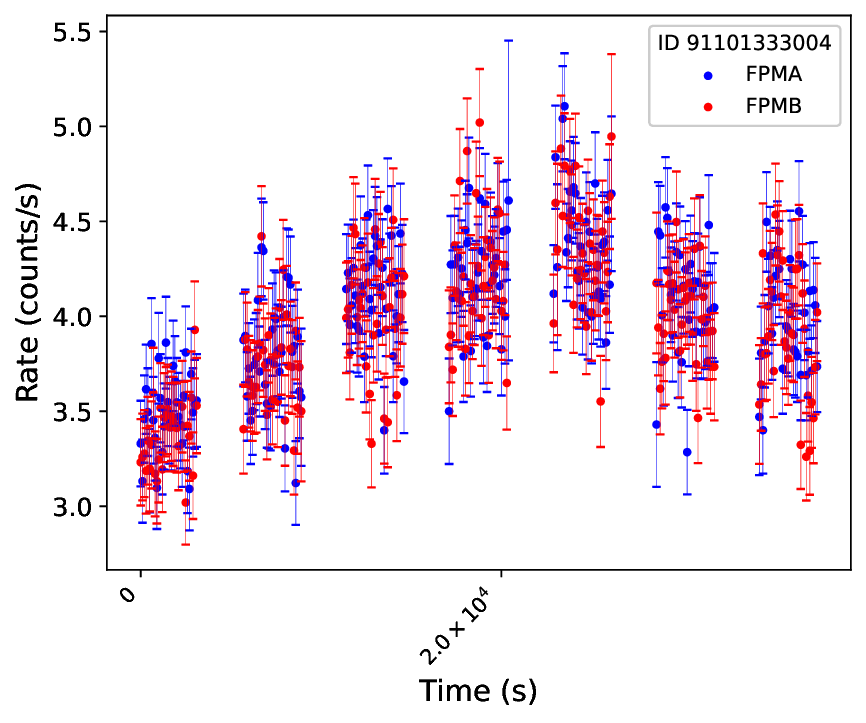}
			\caption{Light curves corresponding to the \textit{NuSTAR} FPMA/FPMB observation of the source SAX J1808 corresponding to the persistent emission of obs IDs 80701312002(\textit{left}), 91101333002(\textit{centre}), and 91101333004(\textit{right}) with bin size 500s, 100s and 100s, respectively in the 3-79 keV energy regime.}
			\label{fig:entire_lc}
		\end{figure*}

		\begin{figure}
			\centering
			\includegraphics[width=0.8\columnwidth]{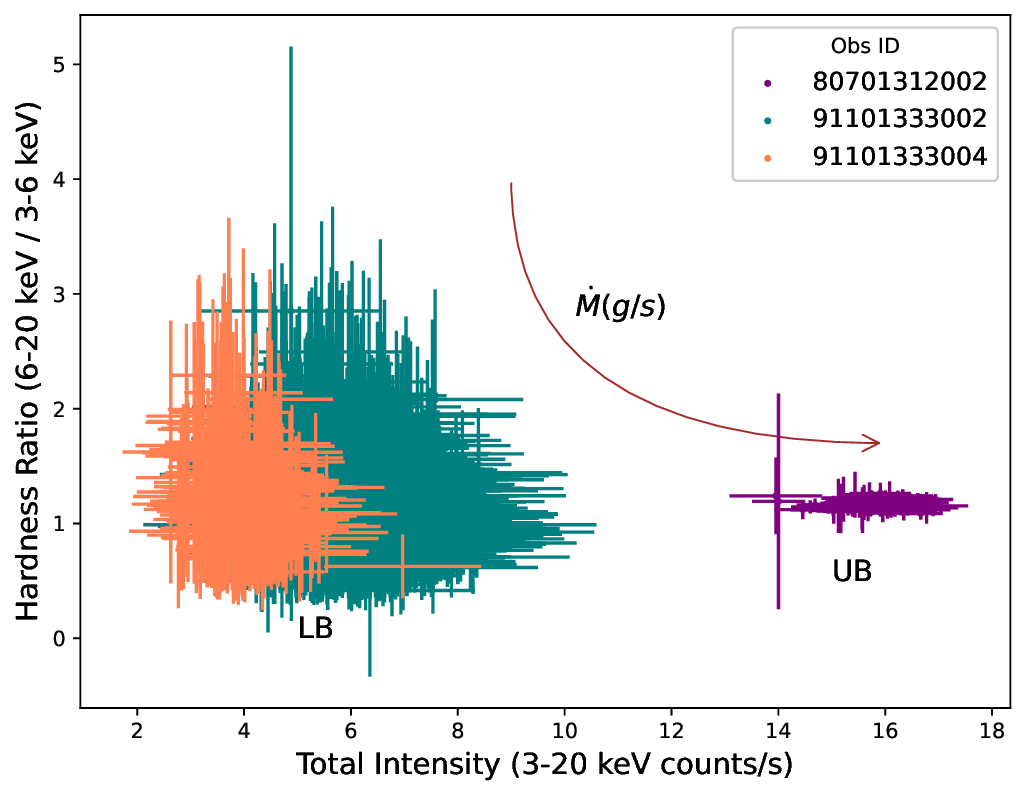}
			\caption{Hardness-intensity diagram for the three \textit{NuSTAR}  observations of 2022 August, 2025 August and September shown in violet, cyan and coral red, respectively. The dashed brown arrow indicates the increase in mass accretion rate \text{$\dot{M}$} (in g/s) across the three observations (Table \ref{tab:mass_acc}).}
			\label{fig:HI_diag}
		\end{figure}
		
		\begin{figure*}
			\centering	
			\begin{minipage}{0.6\textwidth}
				\centering
				\includegraphics[width=\linewidth]{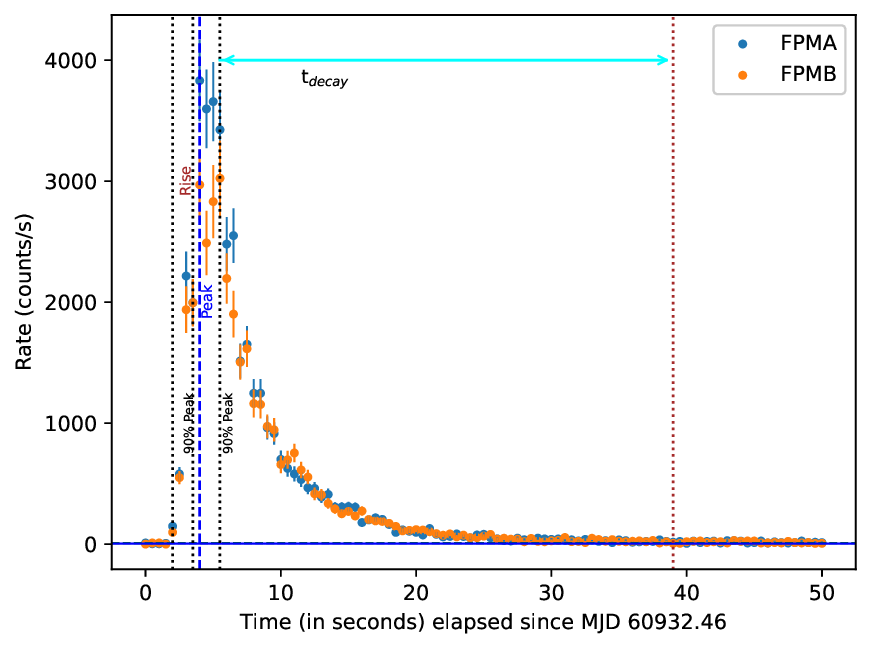}
			\end{minipage}
			\hfill
			\begin{minipage}{0.35\textwidth}
				\centering
				\includegraphics[width=\linewidth]{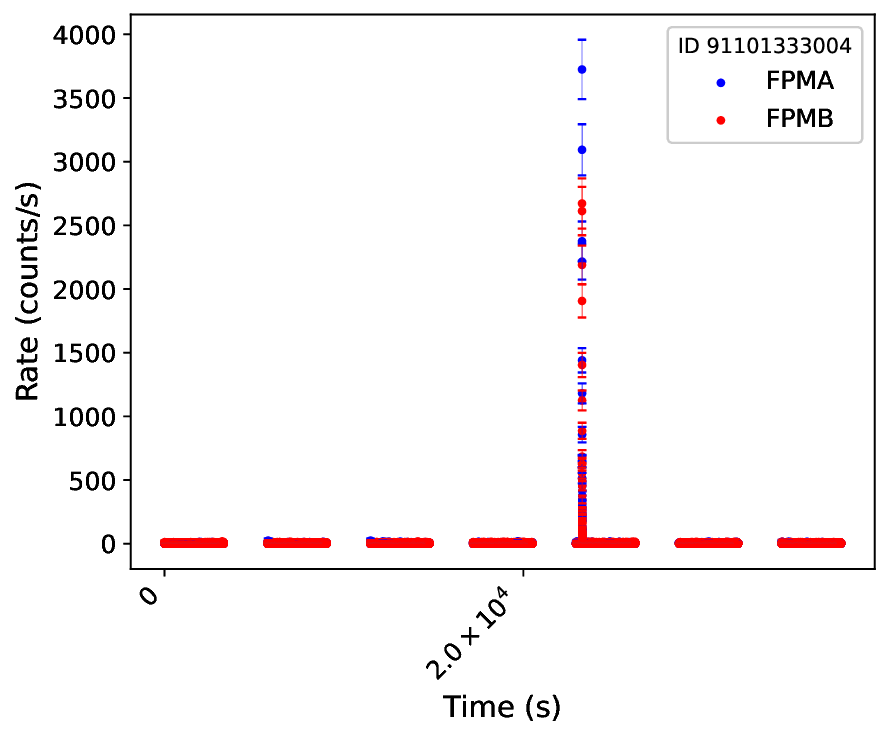}
				
				\vspace{0.2cm}
				
				\includegraphics[width=\linewidth]{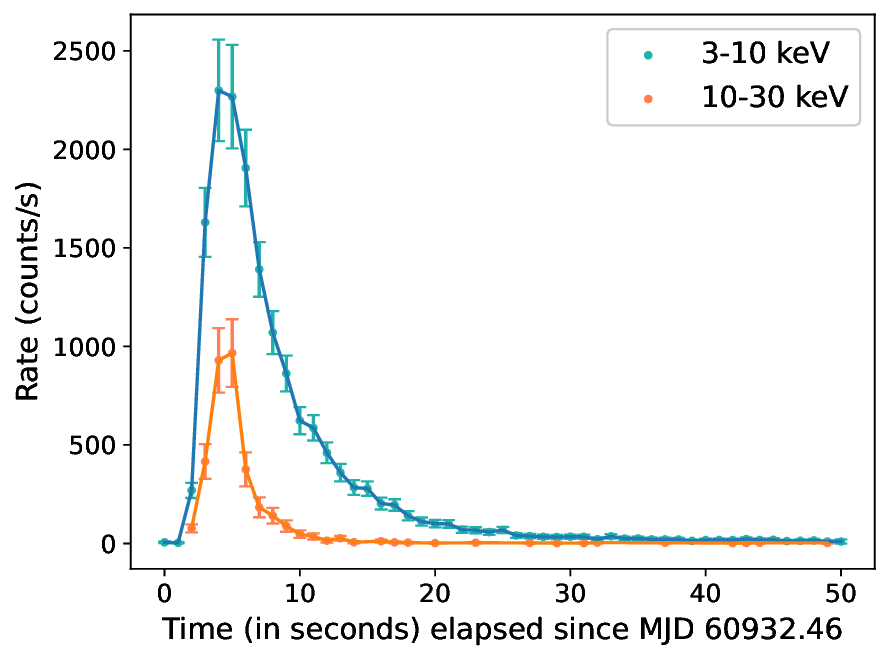}
			\end{minipage}
			\caption{Light curve corresponding to the burst emission \text{$\sim$50 s} duration from 23,134 s to 23,184 s of the \textit{NuSTAR} FPMA/FPMB observation for the source SAX J1808 corresponding to the observation ID 91101333004 \textit{left panel:} with a bin size of 0.5 s in the 3-79 keV energy region. \textit{Upper right:} entire exposure with a bin size of 73.72 s in the 3-79 keV energy regime. \textit{Lower right:} burst exposure observed in 3-10 keV and 10-30 keV energy range shown by blue and orange colors, respectively.}
			\label{fig:sep2025}
		\end{figure*}

		The \textit{MAXI} light curve of the source SAX J1808 with a one-day bin size for the duration of August 2025 to December 2025 in the 2-6 keV energy range is shown in the left panel of Fig. \ref{fig:maxi_data} \citep{2009PASJ...61..999M}. The 2025 outburst was first recorded on 2025 August 06 \citep{2025ATel17323....1R}. On 2025 September 02, \textit{IXPE} detected one thermonuclear burst during the reflaring phase of the outburst \citep{2025ATel17369....1B}. It is clear that the 2025 \textit{NuSTAR} observations reported in this work were carried out in the decay phase of the outburst. The right panel of Fig. \ref{fig:maxi_data} shows the \textit{Swift BAT} light curve for August 2025 to December 2025 with a one-day bin size within 15-50 keV energy range \citep{Ajello_2008}.
		
		The complete \textit{NuSTAR} FPMA/FPMB light curves for the observations 1, 2, and 3 (persistent emission) are shown in the left, middle, and right panels of Fig. \ref{fig:entire_lc}, respectively, in the energy range 3-79 keV. A thermonuclear burst of $\sim50$ s duration has been observed around 23,134s from the onset of observation 3. A time interval of $\sim100$ s is excluded from the \textit{NuSTAR} FPMA/FPMB data for observation 3 to obtain the resulting persistent emission. The average persistent count rate during observations 1, 2, and 3 is $\sim15-18$, $\sim5-7$ counts s$^{-1}$, and $\sim3-5$ counts s$^{-1}$, respectively. As expected, the source was detected in a higher count rate around the peak of the 2022 outburst compared to the decay phase of the 2025 outburst.

		To further investigate the above-mentioned fact, we have extracted the H-I diagram (HID) of the source, as the spectral state of the source can be inferred from it. The hardness-intensity (H-I) diagram represents the distribution of data points with their hardness (ratio of hard energy to soft energy) to their corresponding intensity ($\sim$count rate; \cite{fender2004towards}). We defined the hardness ratio (HR) as the ratio of the source count rate in the 6-20 keV band to that in the 3-6 keV band, and the 3-20 keV count rate was used to estimate the total intensity. The H-I diagrams for the 2022 August, 2025 August, and September observations are shown in violet, cyan, and coral colors, respectively, in Figure \ref{fig:HI_diag}. During 2022 August the source occupies the high-intensity end of the diagram. The stretch of hardness band is narrow that suggests a small change in the HR during the observation, inferring no significant change in HR. During 2025 observations, the source is detected in a low-intensity region towards the left end of the diagram corresponding to the lowest count rate. A change in the mass accretion rate is involved for such observed behavior, which is studied in the latter part of this work.
		
		
		The light curve of the burst emission, detected during Obs 3, for an exposure of $\sim$ 50 s in the 3–79 keV band is shown in the left panel of Fig. \ref{fig:sep2025}. Following \citealt{galloway2008thermonuclear}, we have considered the time duration for the peak of the burst as the total time for which the count rate remains above 90\% of the peak count rate. Rise time for the burst is considered as the time taken to increase from the initial level to the level of 90\% of the peak count rate. Decay time for the burst is considered to be the time taken for the count rate to fall below the initial level. During the peak of the burst, the count increases to $\sim$ 4000 counts s$^{-1}$. The count rate again reaches around the persistent level of $\sim$ 5 counts s$^{-1}$ after 37 s from the peak of the burst. All time scales, including the peak of the burst, have been marked with dotted vertical lines in the left panel of Fig. \ref{fig:sep2025}.\\
		We have also extracted the energy-resolved light curve profile of the burst. The burst light curves in the soft (3-10 keV) and hard (10-30 keV) energy bands are shown in the lower right panel of Fig. \ref{fig:sep2025}. The burst shows a rapid rise to a high count rate and decays slowly in the soft energy band (3-10 keV). This is a characteristic of a fast rise and exponential decay (FRED) profile for thermonuclear bursts. Whereas within the 10-30 keV energy region, the light curve of the burst shows a slower rise and faster decay. We observed a hard X-ray deficit during the burst, indicative of burst-induced cooling of the corona. Similar hard X-ray decrements during bursts have been observed earlier in other neutron star LMXBs (see, e.g., \cite{ji2014hard, sanchez2020burst}). However, to establish the fact, one has to study the effect of the burst photons on the Comptonizing medium.

		\section{Spectral Analysis}
		\label{sec:spec_analys}
		We performed spectral analysis of the persistent and burst emission using \texttt{XSPEC v}12.15.1 (\citealt{1996ASPC..101...17A}). To account for interstellar absorption, we applied the neutral hydrogen ($N_{\rm H}$) component of the \texttt{TBabs} model, adopting \texttt{wilm} abundances \citep{2000ApJ...542..914W} and \texttt{vern} photoelectric cross-section \citep{1996ApJ...465..487V}.  The neutral hydrogen column depth density is kept constant at $n_{\rm H}\sim0.21\times10^{22}$ $cm^{-2}$ \citep{1985SSRv...40..287V, 1990ARA&A..28..215D, 2026ApJ...999..133S}. We assumed a source distance of $\sim$3.5 kpc \citep{galloway2006helium}. Moreover, for visualisation of the spectral plots using \textsc{QDP/PLT} package \citep{tennant1991qdp} within \textsc{XSPEC}, we have rebinned the data points by factors of 100, 50, and 30 for Obs 1, Obs 2 and 3, and Obs 5, respectively, to improve the clarity of presentation (see Fig. \ref{fig:obsA}, \ref{fig:obsB}, \ref{fig:obsC} and \ref{fig:joint}).
		
		\subsection{Persistent spectral analysis}
		\label{subsec:pers_time}
		
		The full exposure of Obs 1, 2, and 3 (persistent) from both the \textit{NuSTAR} FPMA and FPMB detector were fit simultaneously in the energy band 3-79 keV (Obs 1) and 3-50 keV (for Obs 2 and 3), where the source counts are well above the background counts. A constant multiplication factor was included to account for the cross-calibration of FPMA and FPMB instruments. All the parameters are ascertained within 1$\sigma$ uncertainty.
		\begin{table*}
			\centering
			\caption{Best-fit values for the \textit{NuSTAR} and \textit{NICER} observations for the continuum fit corresponding to Model 1: \texttt{const*TBabs*(po+diskbb)} and Model 2:  \texttt{const*TBabs*thcomp*diskbb} in the energy range 3-79 keV for Obs 1, 3-50 keV for both Obs 2 and 3 and in the energy regime 1.2-8.0 keV for Obs 5. All uncertainties are quoted at the 1$\sigma$ confidence level.}
			\begin{tabular}{|c c c |c c c c|}
				\hline
				\parbox{1cm}{Model\\ Component} & Parameter & Unit & \multicolumn{4}{|c|}{Value} \\
				\hline
				\multicolumn{3}{|c|}{Observations}  & \parbox{2.2cm}{Joint fit (1 \& 5)} & 1 & 2 & 3 \\
				\hline
				\multicolumn{7}{|c|}{Model 1: \texttt{const*TBabs*(po+diskbb)}}\\
				\hline
				&&&&&&\\
				\textsc{const} & Factor & \parbox{2cm}{\textit{NuSTAR}(FPMB) \\ \textit{NICER}} & \parbox{1.5cm}{$0.98\pm0.01$ \\ $0.94\pm0.01$} & \parbox{1.5cm}{$0.98\pm0.01$ \\ } & \parbox{1.5cm}{$1.02\pm0.01$ \\ } & \parbox{1.5cm}{$1.01\pm0.01$ \\ } \\
				&&&&&&\\
				\textsc{TBabs} & $N_H$ & 10$^{22}$ cm$^{-2}$ & 0.25 $\pm$ 0.01 & 0.21 (f) & 0.21 (f) & 0.21 (f) \\
				\textsc{Power law} & $\Gamma$ & & 1.93$\pm$ 0.01 & 1.93 $\pm$ 0.01 & 1.99 $\pm$ 0.01 & 2.03 $\pm$ 0.01 \\
				& norm & $10^{-2}$ & 17.1 $\pm$ 0.1 & 16.7 $\pm$ 0.1 & 7.2 $\pm$ 0.2 & 4.7 $\pm$ 0.1 \\
				\textsc{Diskbb} & T$_{\rm{in}}$ & keV & 0.84$\pm$0.01 & 0.83 $\pm$ 0.03 & 0.72 $\pm$ 0.05 & 0.62$\pm$ 0.06 \\
				& norm & & 13 $\pm$ 1 & 14$_{-1}^{+2}$ & 14$_{-5}^{+8}$ & 22 $_{-10}^{+23}$ \\
				\textsc{cflux} & F$_{\rm po}$ & $10^{-10}$erg cm$^{-2}$s$^{-1}$& 12.92 $\pm$ 0.01 & 10.47$\pm$0.01 & 3.69$\pm$0.01 & 2.26$\pm$0.01 \\
				& F$_{\rm diskbb}$ & $10^{-10}$erg cm$^{-2}$s$^{-1}$ & 1.21 $\pm$ 0.01 & 0.24$\pm$0.01 & 0.09$\pm$0.01 & 0.04$\pm$0.01 \\
				& F$_{\rm Total}$ & $10^{-10}$erg cm$^{-2}$s$^{-1}$& 14.13 $\pm$ 0.02 & 10.70$\pm$0.01 & 3.78$\pm$0.01 & 2.32$\pm$0.01 \\
				\hline
				& $\chi^2 / dof$ & & 2340/1756 & 2151 / 1644 & 684/655 & 885 / 882 \\
				\hline
				\multicolumn{7}{|c|}{Model 2: \texttt{const*TBabs*thComp*diskbb}}\\
				\hline
				\textsc{thcomp} & $\Gamma_{\rm \tau}$ &  & 1.90$\pm$0.01 & $\lesssim$1.91 & 1.95$\pm$0.01 & 1.99$\pm$0.01 \\
				& kT $_{\rm e}$ & keV & 43$\pm$5& 46$\pm$1 & 28$_{-4}^{+10}$  & 86$_{-26}^{+6}$ \\
				& \parbox{3cm}{Covering fraction (\texttt{cov$\_$frac})} & & 0.74$\pm$0.02 & 0.89$\pm$ 0.01 & 0.78$_{-0.04}^{+0.07}$ & $\lesssim$0.79 \\
				& Redshift (z) & & 0 (f) & 0 (f) & 0 (f) & 0 (f) \\ 
				\textsc{diskbb} & T$_{in}$ & keV & 0.81$\pm$0.01 & 0.76 $\pm$ 0.01 & 0.72$\pm$ 0.03 & 0.63$_{-0.05}^{+0.06}$ \\ 
				& norm &  & 58$\pm$3 & 156$\pm$1 &  81$_{-19}^{+15}$ & 147$_{-58}^{+50}$ \\
				\textsc{cflux} & F$_{\rm thcomp*diskbb}^{\rm a}$ & $10^{-10}$erg cm$^{-2}$s$^{-1}$ & 14.36$\pm$0.02 & 11.10$\pm$0.10 & 3.78$\pm$0.01 & 2.38$\pm$0.01 \\
				\hline
				& $\chi^2$/dof & & 2239/1755 & 2070 /1643 & 670/654 & 883/881 \\
				\hline 
				\multicolumn{7}{l}{\parbox{16cm}{\textit{Note:} All the unabsorbed flux are estimated for the 3-79 keV energy band using \textsc{cflux} model in \textsc{xspec}. }}
			\end{tabular}
			\label{tab:mod1,2}
		\end{table*}
		
		Initially, to ascertain the thermal emission from the disc, we utilise the multicolor disc model \texttt{diskbb} (\citealt{1984PASJ...36..741M}; \citealt{1986ApJ...308..635M}). To estimate the Comptonized/non-thermal emission from the corona, we added a \texttt{power law} model. So, we have used \texttt{const*TBabs*(po+diskbb)} (Model 1). For Obs 1, 2, and 3, Model 1 resulted in $\chi^2/dof$ = 2151/1644, 684/655, and 885/882, respectively. The corresponding best-fit values are shown in Table \ref{tab:mod1,2}. From the continuum fit, we observed excess around 5-8 keV, indicating the presence of broad Fe emission, and within the 15-30 keV range, known as the Compton hump. The power-law flux was estimated by incorporating the \texttt{cflux} component with the \texttt{power law} component in \textsc{xspec}. The power law flux contains roughly $\sim95\%$ of the total flux for the three observations. The corresponding power-law flux values (Table \ref{tab:mod1,2}) are an order of magnitude higher than the \texttt{diskbb} flux, indicating that the Comptonized emission dominates over the disc emission in all three observations. Overall, it indicates that the three observations are at a hard spectral state.
		
		\begin{figure*}[t]
			\centering
			\includegraphics[width=0.48 \textwidth]{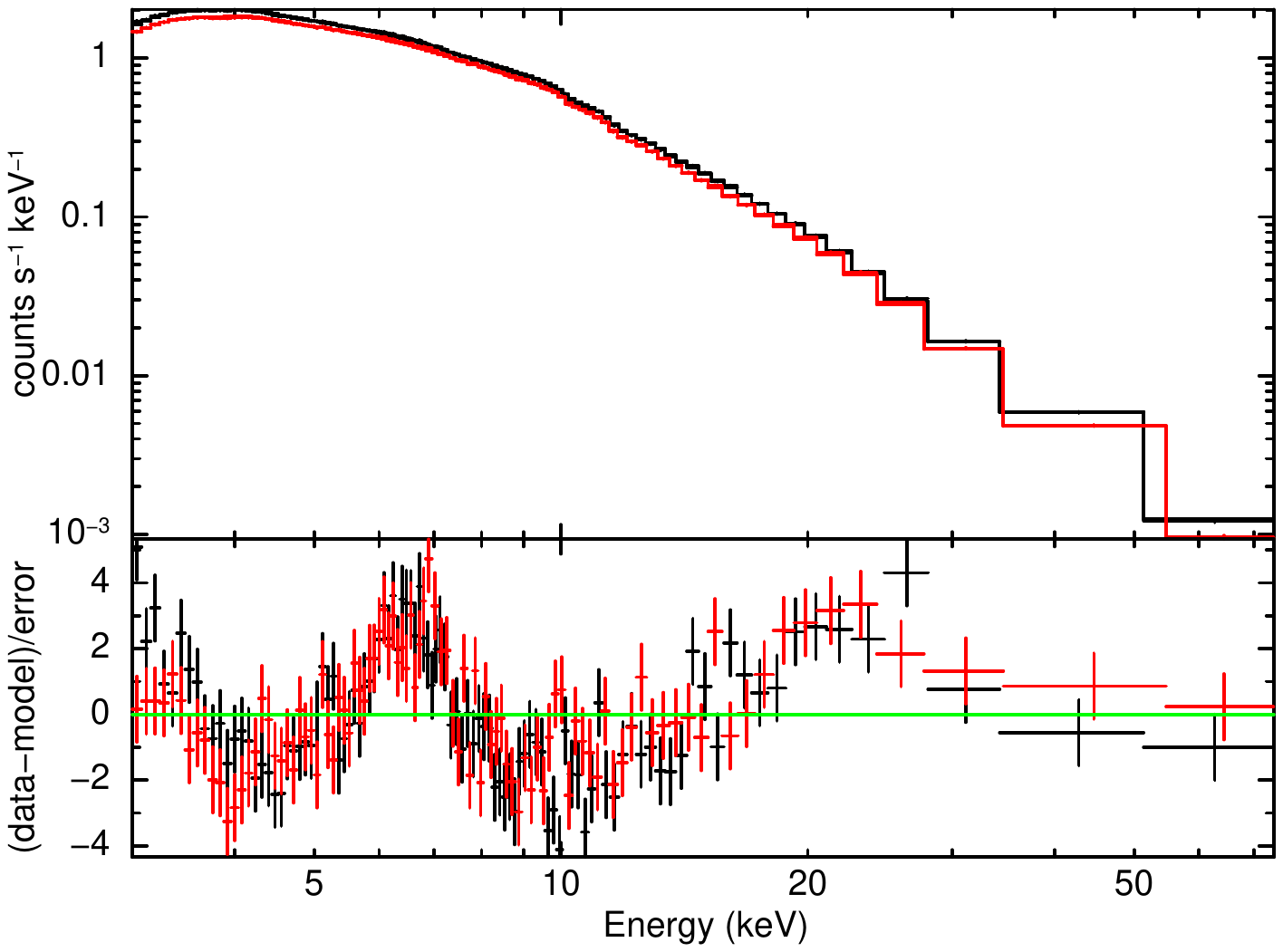}
			\includegraphics[width=0.48 \textwidth]{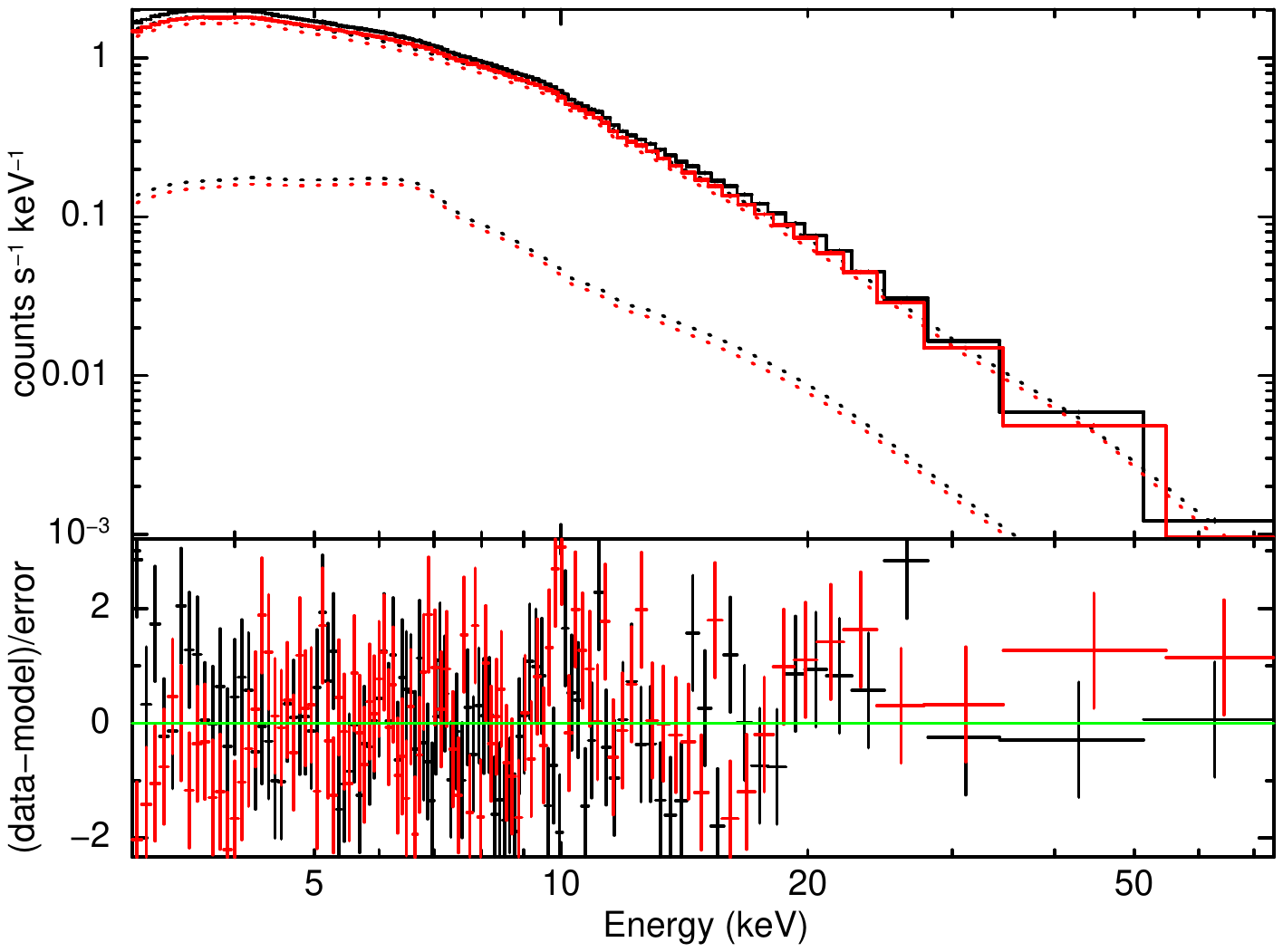}\\
			\caption{Spectral plot for the \textit{NuSTAR} FPMA (\textit{black})/FPMB (\textit{red}) observation of the source SAX J1808 corresponding to the Obs 1 in the energy range 3-79 keV for the - \textit{left Panel:} continuum emission with the Model 2: \texttt{const*TBabs*(thComp*diskbb)} \textit{right panel:} unfolded spectra corresponding to the best-fit Model 3: \texttt{const*TBabs*(thComp*diskbb+relxillCP)}. The lower panels of plots show the residuals in 1 $\sigma$ uncertainty. Data have been rebinned by a factor of 100 for visual purposes.}
			\label{fig:obsA}
		\end{figure*}
		
		\begin{figure*}
			\centering
			\includegraphics[width=0.48 \textwidth]{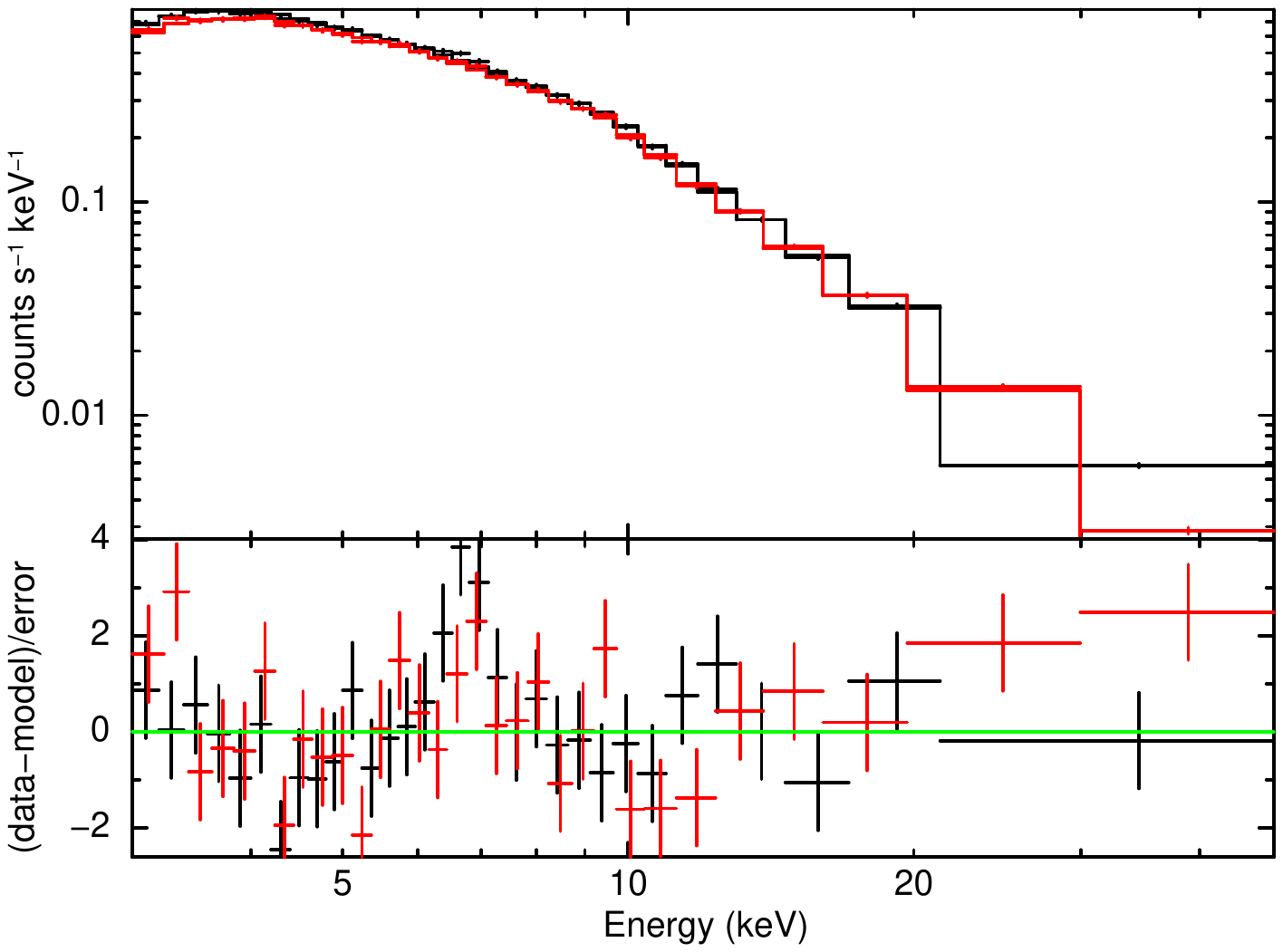}
			\includegraphics[width=0.48 \textwidth]{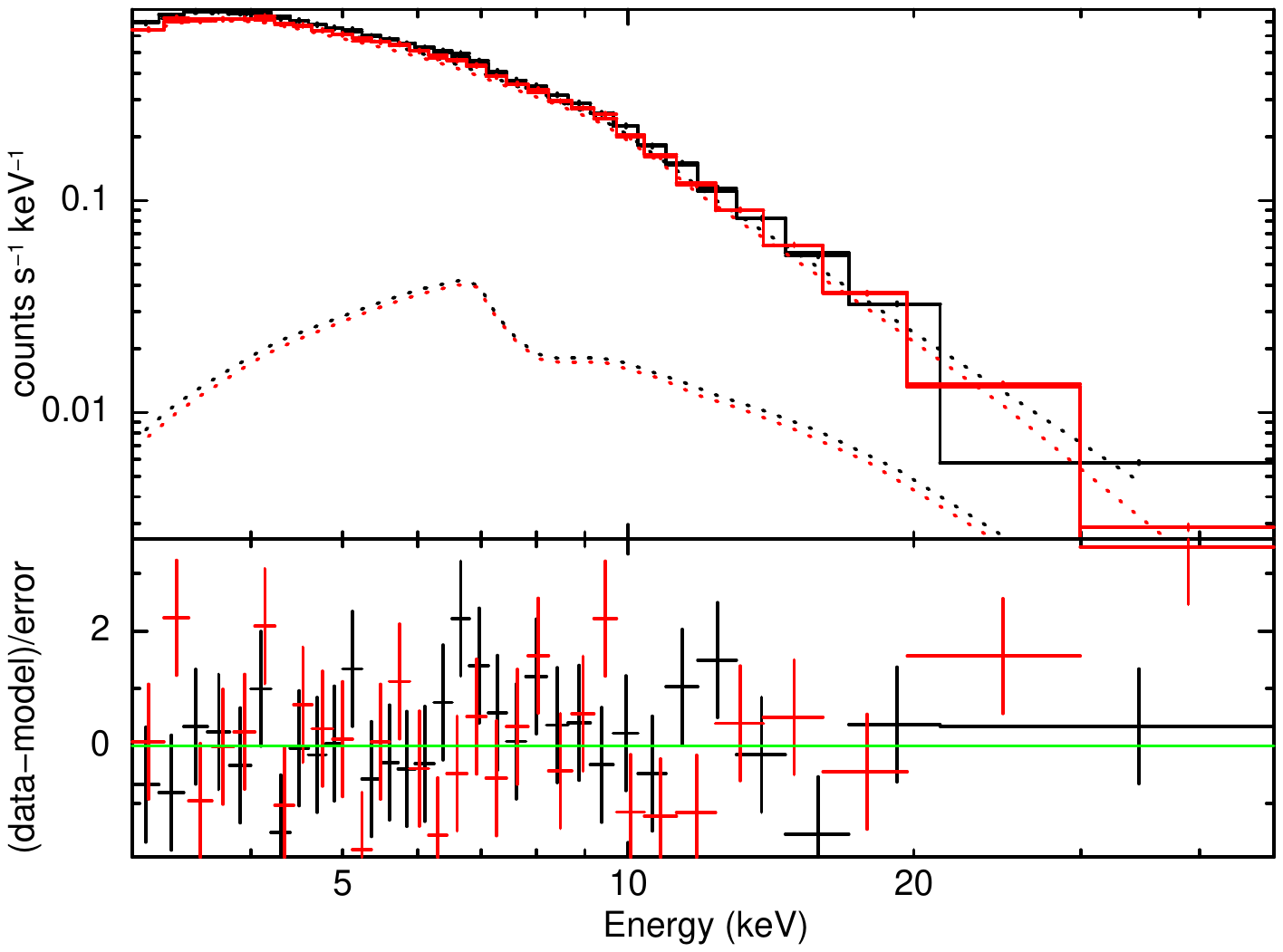}\\
			\caption{Spectral plot for the \textit{NuSTAR} FPMA (\textit{black})/FPMB (\textit{red}) observation of the source SAX J1808 corresponding to Obs 2 in the energy range 3-50 keV for the - \textit{left Panel:} continuum emission with the Model 2: \texttt{const*TBabs*(thComp*diskbb)}, \textit{right panel:} unfolded spectra corresponding to the best-fit Model 3: \texttt{const*TBabs*(thComp*diskbb+relxillCP)}. The lower panels of the figures show the residuals in 1 $\sigma$ uncertainty. Data have been rebinned by a factor of 50 for visual purposes.}
			\label{fig:obsB}		
		\end{figure*}
		
		\begin{figure*}
			\centering
			\includegraphics[width=0.48 \textwidth]{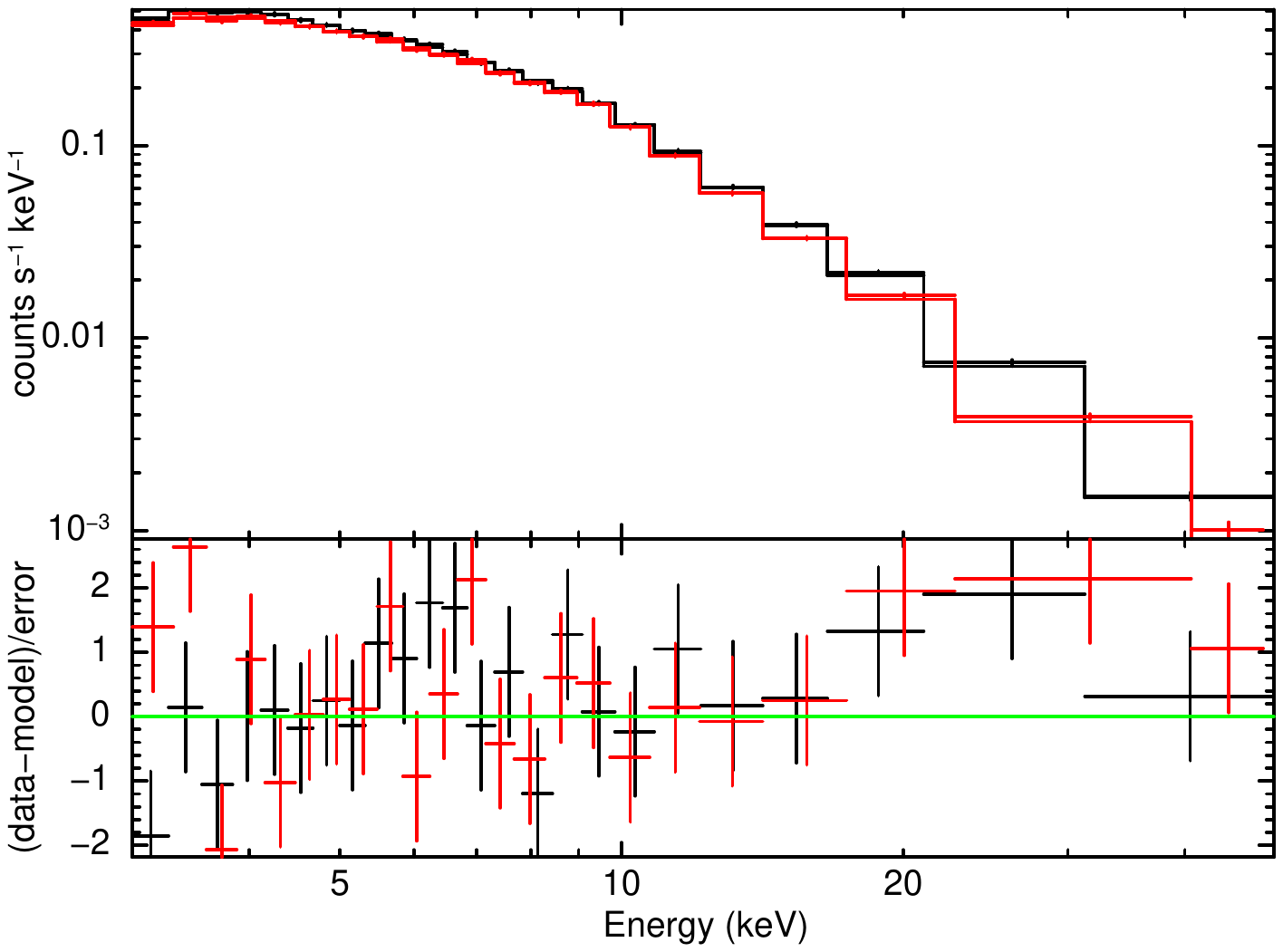}
			\includegraphics[width=0.48 \textwidth]{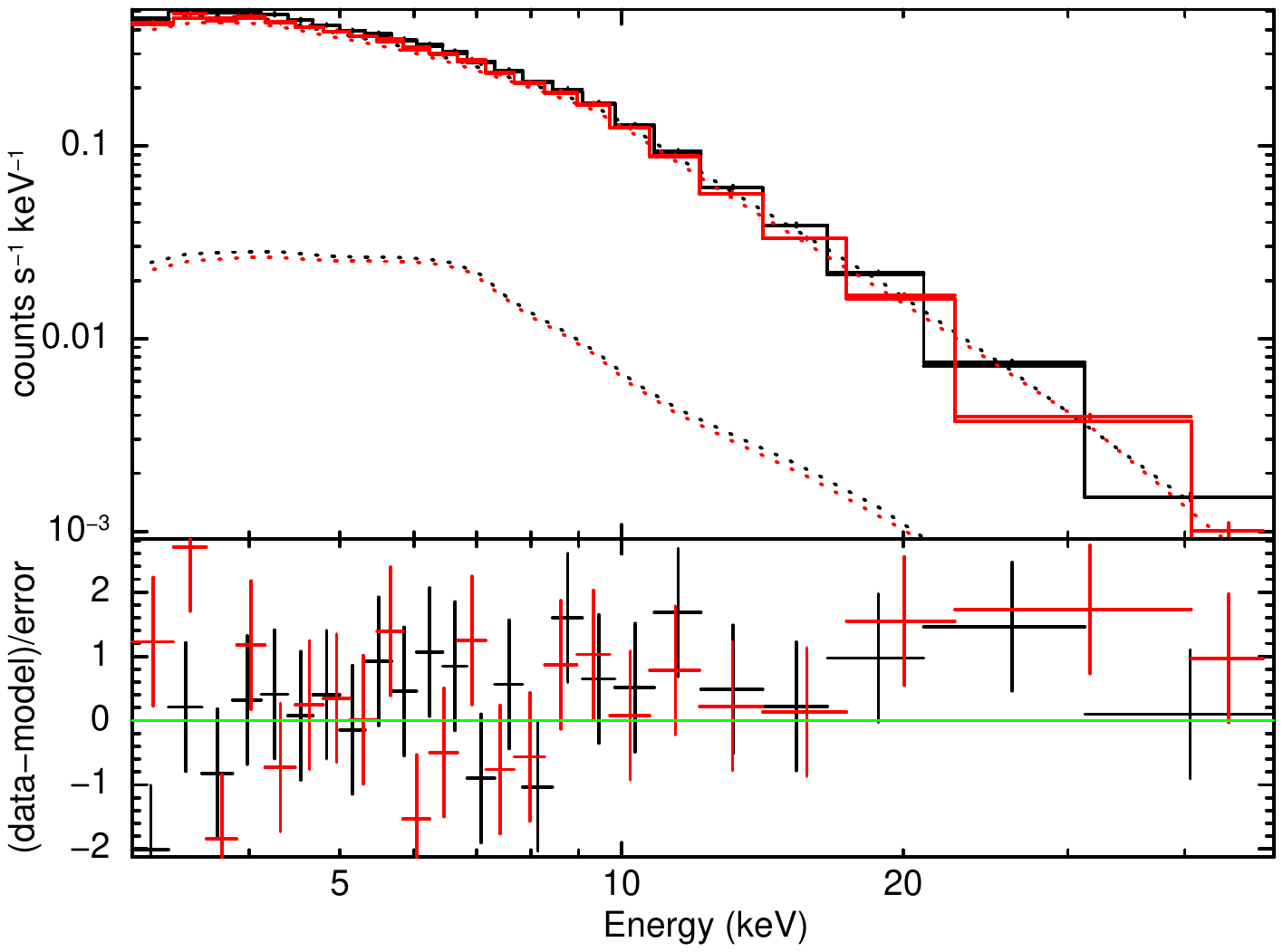}\\
			\caption{Spectral plot for the  persistent emission of \textit{NuSTAR} FPMA (\textit{black})/FPMB (\textit{red}) observation of the source SAX J1808 corresponding to the Obs 3 in the energy range 3-50 keV for the \textit{Left Panel:} continuum emission with the Model 2: \texttt{const*TBabs*(thComp*diskbb)}, \textit{right panel:} unfolded spectra corresponding to the best-fit Model 3: \texttt{const*TBabs*(thComp*diskbb+relxillCP)}. The lower panels of the figures show the residuals in 1 $\sigma$ uncertainty. Data have been rebinned by a factor of 50 for visual purposes.}
			\label{fig:obsC}
		\end{figure*}
		
		To account for the excess in the observed continuum spectra in the 5-8 keV range, we added the \texttt{gaussian} to Model 1 for Obs 1, 2, and 3. The iron line emission energy was estimated to be 6.4 keV, 6.7 keV, and 6.8 keV with equivalent line width (EW) $\sim$53, $\sim$43, and $\sim$20 eV, respectively. To analyse the Comptonised emission from the corona more accurately, we replaced the \texttt{power law} component in Model 1 with the convolution model \texttt{thComp} \citep{1996MNRAS.283..193Z, 2019MNRAS.485.2942N, 2020MNRAS.492.5234Z}. The thermal comptonization model (\texttt{thComp}) describes the continuum Comptonisation emission much better than a cutoff power law model and is an updated version of \texttt{nthComp} that aligns more closely with the actual Monte Carlo spectra from Comptonization than \texttt{nthComp} does. The \texttt{thComp} model estimates the Comptonization distribution of seed photons from the disc as well as the neutron star surface. The component covering fraction (\texttt{cov$\_$frac}) signifies the portion of seed photons that has been taken up from the disc into the corona. These photons get upscattered from the corona. We used the Model 2 (\texttt{const*TBabs*(thComp*diskbb)}) for the analysis of the continuum. The Model 2 (\texttt{const*TBabs*(thComp*diskbb)}) provided $\chi^2 / dof$ = 2070/1643, 670/654, and 883/881 ($\Delta \chi^2$ = -81 for 1 $dof$, -14 for 1 $dof$, and -2 for 1 $dof$) corresponding to Obs 1, 2, and 3, respectively. The best-fit values corresponding to Model 2 are provided in Table \ref{tab:mod1,2}. The corresponding spectral plots with 1$\sigma$ uncertainty residuals, are shown in the left panel of Fig. \ref{fig:obsA}, \ref{fig:obsB} and \ref{fig:obsC}. Data have been rebinned by a factor of 100 for Obs 1, and 50 for both Obs 2 and 3. The electron temperature (kT$_{\rm e}$) obtained in Model 2 estimated $\sim$46, $\sim$28, and $\sim$86 keV for Obs 1, 2, and 3, respectively. The kT$_{\rm e}$ values indicate the presence of a hot Comptonizing corona. In addition, Model 2 indicates a high covering fraction (\texttt{cov$\_$frac}), around 80-90$\%$, when the seed photons originate from the accretion disc. The presence of broad Fe emission line and Compton hump suggest that the photons get reflected from the accretion disc, signifying the disc reflection, which has been observed earlier for many accreting millisecond X-ray pulsars \citep{Steiner2017, Garcia2020, maha24, maha25}.
		
		\begin{table*}
			\centering
			\caption{Best-fit results for the parameters of the components corresponding to the \textit{NuSTAR} and \textit{NICER} observations using Model 3: \texttt{const*TBabs*(thComp*diskbb+relxillCP)} in the energy range 3-79 keV for Obs 1, within 3-50 keV for both Obs 2 and 3 and in the energy regime 1.2-8.0 keV for Obs 5. All uncertainties are quoted at the 1$\sigma$ confidence level.}
			\begin{tabular}{| c c c | c c c c |}
				
				\hline
				\parbox{3cm}{Model Component} & Parameter & Instrument/Unit & \multicolumn{4}{|c|}{Value} \\
				\hline
				\multicolumn{3}{|c|}{Observation} &  \parbox{2.2cm}{Joint fit (1 \& 5)} & 1 & 2 & 3 \\
				\hline
				&&&&&&\\
				\textsc{const} & Factor & \parbox{2cm}{\textit{NuSTAR}(FPMB) \\ \textit{NICER}} & \parbox{1.5cm}{$0.98\pm0.01$ \\ $0.94\pm0.01$} & \parbox{1.5cm}{$0.98\pm0.01$ \\ } & \parbox{1.5cm}{$1.01\pm0.01$ \\ } & \parbox{1.5cm}{$1.01\pm0.01$ \\ } \\
				&&&&&&\\
				\textsc{TBabs} & $N_H$ & 10$^{22}$ cm$^{-2}$ & 0.15$\pm$0.04 & 0.21 (f) & 0.21 (f) & 0.21 (f) \\
				\textsc{relxillCp} & Incl & deg($^\circ$) & 48$_{-9}^{+8}$ & $37_{-3}^{+2}$ & 45$_{-4}^{+15}$ & 39$_{-17}^{+7}$ \\
				& R$_{\rm in}$ & R$_{\rm ISCO}$ & 4.6$_{-2.4}^{+2.1}$ & $\lesssim$1.5 & $\lesssim$7.0 & 4.8$_{-2.9}^{+3.3}$ \\
				& $\Gamma$ &  &  1.94$\pm$0.01 & $1.93\pm0.02$ &2.05$_{-0.03}^{+0.02}$ & 1.99$\pm$0.02 \\
				& log $\xi$ & & 2.45$_{-0.08}^{+0.21}$ & $2.83\pm0.05$ & $\lesssim$1.30 & $3.22\pm0.50$ \\
				& log N & cm$^{-3}$ & 18.5$_{-1.7}^{+0.6}$ &18 (f) & 18 (f) & 18 (f) \\
				& A$_{\rm Fe}$ & & 1.00$_{-0.10}^{+0.52}$ & 1 (f) & 1 (f) & 1 (f) \\
				& k$T_{\rm e}$ & keV & $\gtrsim$54 & 83$_{-35}^{+3}$ & 85$_{-37}^{+10}$ & 100$_{-51}^{+10}$  \\
				& \centering \parbox{2cm}{Reflection fraction} & & -1.00(f) & -1.00 (f) & -1.00 (f) & -1.00 (f) \\
				& norm & $10^{-4}$ & 11.4$_{-3.8}^{+3.1}$ & $19.6_{-2.4}^{+0.8}$ & $1.0_{-0.3}^{+0.1}$ & $3.3_{-1.3}^{+5.1}$ \\
				\textsc{diskbb} & T$_{in}$ & keV & 0.80$\pm$0.01 & 0.69$_{-0.01}^{+0.02}$ & 0.63$_{-0.03}^{+0.02}$ & 0.61$_{-0.04}^{+0.03}$ \\
				& norm &  & 60$_{-3}^{+5}$ & 219$_{-9}^{+12}$ & 157$_{-46}^{+52}$ & 142$_{-49}^{+52}$ \\
				\textsc{thcomp} & $\Gamma_{\rm \tau}$ & & \multicolumn{4}{c|}{= $\Gamma$ of \textsc{relxillCp}}  \\
				& \centering \parbox{2cm}{Covering fraction (\texttt{cov$\_$frac})}  & & 0.87$\pm$0.08 & 0.85$\pm$0.02 & 0.99$_{-0.19}^{+0.01}$ & 0.99$_{-0.29}^{+0.01}$ \\ 
				\textsc{cflux} & F$_{\rm thcomp*diskbb}^{\rm a}$ & $10^{-10}$erg cm$^{-2}$s$^{-1}$ & 13.20$\pm$0.01 & 12.03$\pm$0.01 & 4.22$\pm$0.01 & 1.20$\pm$0.01 \\
				& F$_{\rm relxillCp}$ & $10^{-10}$erg cm$^{-2}$s$^{-1}$ & 1.18$\pm$0.01 & 1.15$\pm$0.01 & 0.46$\pm$0.01 & 0.18$\pm$0.01 \\
				& F$_{\rm Total}$ & $10^{-10}$erg cm$^{-2}$s$^{-1}$ & 14.38$\pm$0.01 & 13.18$\pm$0.01 & 4.68$\pm$0.01 & 3.02$\pm$0.01 \\
				\hline
				& $\chi^2$/dof & & 1785/1749 & 1637/1639 & 633/650 & 873/877 \\
				\hline
				\multicolumn{6}{l}{\parbox{16cm}{\textit{Note:} All the unabsorbed flux are estimated for the 3-79 keV energy band using \textsc{cflux} model in \textsc{xspec}. }}
			\end{tabular}
			\label{tab:spec_analysis}
		\end{table*}
		
		\subsection{Disc Reflection}
		\label{subsec:disc}
		To assess the Comptonized emission reflected from the accretion disc, we used the \texttt{relxillCP} model, an X-ray reflection model that uses the thermal Comptonizing continuum as the primary illuminating source \citep{2014ApJ...782...76G, Dauser_2016}. Allowing a larger number of parameters to vary increases the degeneracy in the model, making it more difficult to constrain the parameters of interest. As there is a large number of parameters involved in the \texttt{relxillCP} model, we fixed those that do not affect the reflection spectrum greatly. This approach enables tighter constraints on the parameters of interest. Our primary focus is to extract information about the accretion disc as the \texttt{thComp} parameter \texttt{cov$\_$frac} indicates $\sim$ 0.8 of the total seed photons are upscattered. We are interested in estimating the inner disc radius and inclination angle of the accretion disc. We have frozen the reflection fraction component at -1 to study only the disc-reflected emission. The dimensionless spin parameter ($a^{*}$) was calculated numerically utilizing the maximum spin parameter for neutron stars in LMXBs ($j_{\rm {max}}$ $\sim$ 0.7; \citealt{lo2011spin}) and the pulsation frequency ($f$) of SAX J1808 ($\sim$401 Hz). The maximum breakup frequency ($f_{\rm k}$) is considered within 1000-1500 Hz. We considered the breakup frequency $\sim$1400 Hz. For slowly rotating stars ($a^* < 0.3$), the spin parameter ($a^{*}$) scales as $a^{*}=j_{\rm max}\frac{f}{f_{\rm k}}$. The value obtained is $a^{*} \simeq 0.2$, which was frozen during the spectral fitting. The photon index ($\Gamma_{\rm \tau}$) of \texttt{thComp} was tied to the photon index ($\Gamma$) of \texttt{relxillCP}. Also, we have tied the electron temperature parameter (k$T_{\rm e}$) of the \texttt{relxillCP} model to the electron temperature parameter of \texttt{thComp} component. The emissivity indices ($q_1$ and $q_2$) are fixed at $\simeq$3, corresponding to the standard value for a centrally radiating source \citep{2019MNRAS.483..767D, ASM_2020, Naik2026}. The break radius ($R_{\rm br}$) is set to be equal to the outer disc radius ($R_{\rm out}$), which is fixed at $\simeq$1000$GM/c^2$. We added the \texttt{relxillCP} component to Model 2 and applied Model 3, \texttt{const*TBabs*(thComp*diskbb+relxillCP)}, to describe the reflected spectrum. We fixed the iron abundance ($A_{\rm {Fe}}$) at the solar abundance level, $A_{\rm Fe}$$\simeq$1, for all three observations. For the three observations, Model 3 provided with $\chi^2 / dof$ = 1638/1639, 633/650, and 873/877 (corresponding to improvements of $\Delta \chi^2 =$ -432 for 4 $dof$, -37 for 4 $dof$, and -10 for 4 $dof$). Applying the F-test to compare Model 3 (reflection) with Model 2 (continuum), we get the $F$-values of $\sim$108, $\sim$10, and $\sim$3 with corresponding null-hypothesis probabilities (p-values) of $\sim$$10^{-81}$, $\sim$1.8$\times10^{-7}$, and $\sim$1.4$\times10^{-2}$ for Obs 1, 2, and 3, respectively. The results indicate that the model expansion is statistically justified, as the $F$-test rejection probability is extremely low for Obs 1 and moderately low but still significant for Obs 2 and 3. The best-fit parameter values obtained from Model 3 are provided in Table \ref{tab:spec_analysis}. The spectral plots for the individual components and residuals corresponding to the best-fit Model 3 are shown in the right panel of Fig. \ref{fig:obsA}, Fig. \ref{fig:obsB}, and Fig. \ref{fig:obsC}. All the spectral plots are shown with 1$\sigma$ uncertainty and the data have been rebinned for visual purposes by a factor of 100 for Obs 1, and 50 for both Obs 2 and 3.\\
		
		We further wanted to examine the presence of the Fe emission line in the 5-8 keV energy regime in the 2022 observation to provide a supplementary basis for the hard X-ray broadband reflection features. Therefore, we considered two \textit{NICER} observations for our analysis, which are very close to the 2022 \textit{NuSTAR} observation, as there were no simultaneous observations between the two satellite missions in 2022 for the source SAX J1808. To fit the \textit{NICER} spectrum, we adopted a similar model combination to that used for the \textit{NuSTAR} spectrum for continuum emission. During the fitting, we have ignored the energy band 0.8-1.2 keV due to known calibration issues in \textit{NICER} detectors \citep{nicer_doc}. We modeled the \textit{NICER} spectrum with the absorbed \texttt{diskbb} plus \texttt{powerlaw} model. This model well describes the continuum emission, with $\chi^2/dof$ = 135/131 and 90/133 for observations 4 and 5, respectively. From both observations, the disc temperature and the power law photon index were found to be around $\sim$0.90 keV and $\sim$1.56, respectively, in agreement with the \textit{NuSTAR} observations.\\
		Most importantly, the \textit{NICER} spectrum also shows prominent residuals around 6-8 keV, indicative of a broad Fe-emission line. Primarily, we fitted those emission lines with Gaussian profiles, which significantly improved the fit. The equivalent widths of the lines are measured to be $\sim0.23$ keV and $\sim0.14$ keV for observations 4 and 5, respectively, with a line centroid energy of $\sim6.42$ keV.
		
		\subsection{Joint spectral analysis}
\begin{figure*}
			\centering
			\includegraphics[width=0.48 \textwidth]{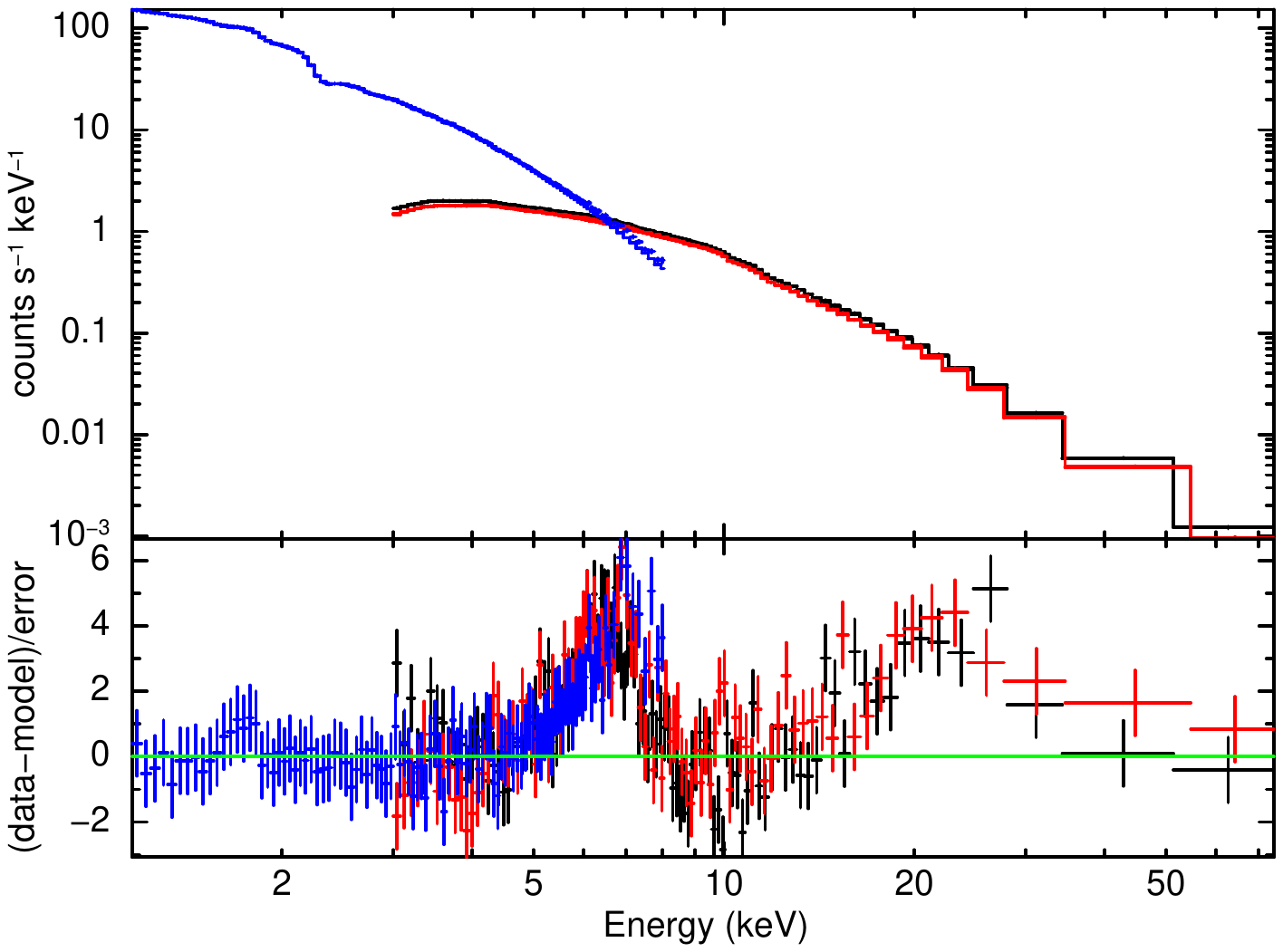}
			\includegraphics[width=0.48 \textwidth]{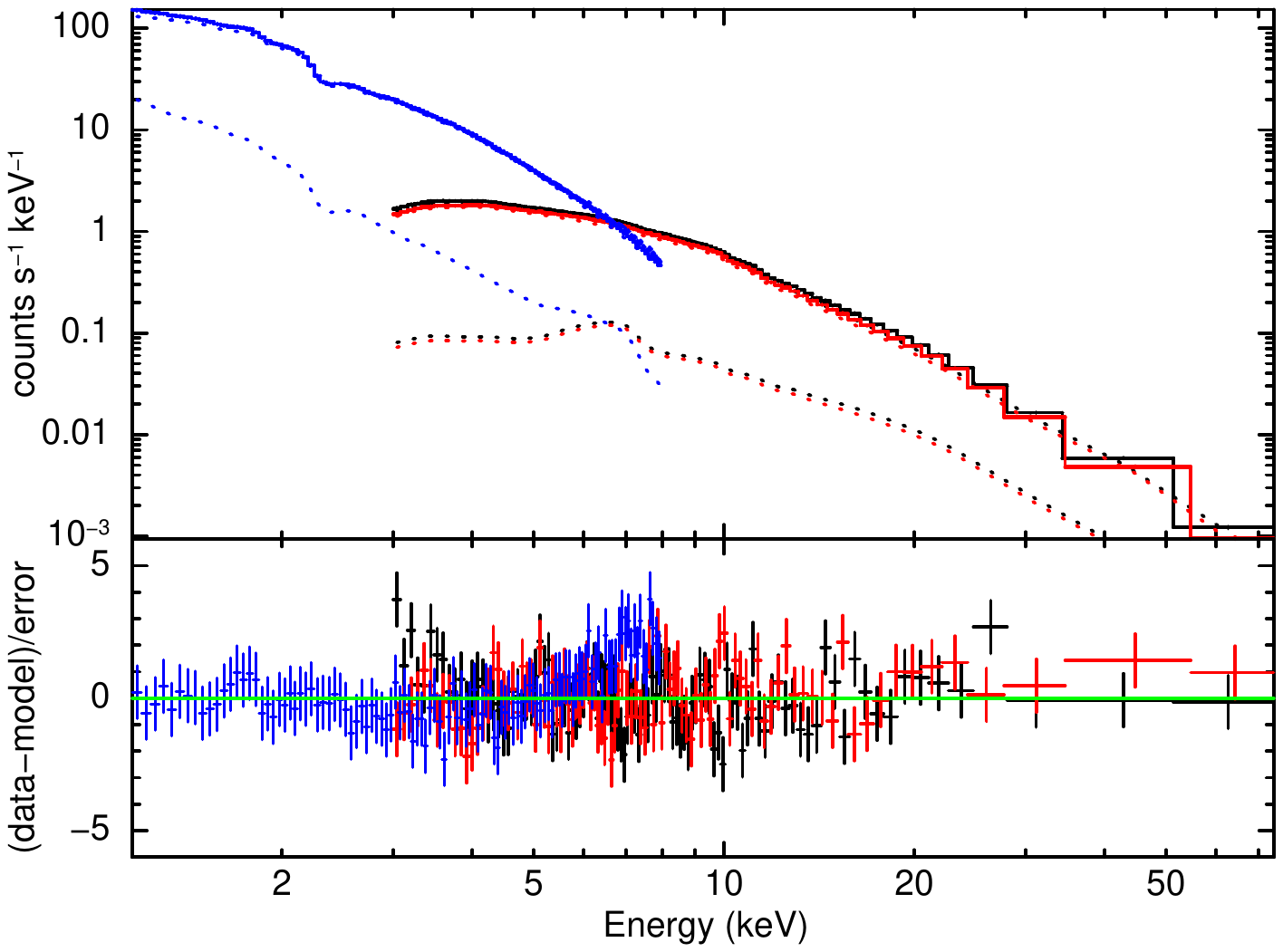}\\
			\caption{Spectral plot for the \textit{NuSTAR} FPMA (\textit{black})/FPMB (\textit{red}) and \textit{NICER}/XTI(\textit{blue}) observation of the source SAX J1808 corresponding to the joint fit of Obs 1 in the energy range 3-79 keV and Obs 5 in the energy band 1.2-8 keV for the - \textit{left Panel:} continuum emission with the Model 2: \texttt{const*TBabs*(thComp*diskbb)} and \textit{right panel:} Model 3: \texttt{const*TBabs*(thComp*diskbb+relxillCp)}. The \textit{NICER} and \textit{NuSTAR} spectra are rebinned by a factor of 30 and 100, respectively, for visual purposes. The lower panels of plots show the residuals in 1$\sigma$ uncertainty.}
			\label{fig:joint}
		\end{figure*}

		As the exposure of \textit{NICER} observation 5 was entirely contained within that of the \textit{NuSTAR} observation 1, we carried out a joint spectral analysis of the two observations. Such an analysis provides more robust constraints on the fundamental spectral parameter with improved spectral fits. We first fitted the spectra using Model 1, \texttt{const*TBabs*(po+diskbb)}, and the corresponding best-fit values are listed in Table \ref{tab:mod1,2}. It provided with $\chi^2/dof=2340/1756$. The fit estimated the value of neutral hydrogen column density to be $N_{\rm H}\sim0.25\times10^{22}$ cm$^{-2}$, consistent with \cite{1990ARA&A..28..215D} and \cite{2026ApJ...999..133S}. The power law index and the inner disc temperature were obtained as $\Gamma\sim1.93$ and k$T_{\rm in}\sim0.84$ keV, respectively, consistent with that obtained for \textit{NuSTAR} Obs 1. To fit the emission observed around 5-8 keV we have utilised the \texttt{Gaussian} model which provided the line emission energy $\sim6.5$ keV with an equivalent width of $\sim$56 eV. The power law model was replaced with \texttt{thcomp} to study the Comptonized emission from the corona. The continuum model provided $\chi^2/dof$=2239/1755 ($\Delta\chi^2=-101$ for 1 $dof$). The photon index and electron temperature for the \texttt{thcomp} model were obtained as $\Gamma_\tau\sim1.90$ and k$T_{\rm e}\sim43$ keV, respectively, which are consistent with the corresponding best-fit values of Obs 1. The best-fit parameters for Model 2, \texttt{const*TBabs*thComp*diskbb}, are provided in Table \ref{tab:mod1,2}. The corresponding spectral fit is shown in the left panel of Figure \ref{fig:joint}, where a broad emission feature around 5-8 keV and a prominent Compton hump around 15-30 keV are observed. We implemented Model 3, \texttt{const*TBabs*(thComp*diskbb+relxillCP)}, to study the reflection emission from the accretion disc. The model corresponding to the reflection analysis provided $\chi^2/dof=1785/1749$ ($\Delta\chi^2=-454$ for 6 $dof$). The best-fit values are tabulated in Table \ref{tab:spec_analysis} and the spectral plot is shown in the right panel of Figure \ref{fig:joint}. The angle of inclination and inner radius of the accretion disc ($R_{\rm in}$) were obtained as $\sim48^\circ$ and $\sim4.6$ $R_{\rm ISCO}$, respectively. The disc electron density parameter was constrained to $\log N\sim18.5$, while the iron abundance was estimated as $A_{\rm Fe}\sim1$, indicating an abundance close to the solar value. The lower limit of the electron temperature (k$T_{\rm e}$) corresponding to Model 3 was obtained as 54 keV. However, it could not be precisely constrained as the upper bound of the 1$\sigma$ uncertainty was not obtained.\\

\begin{figure*}
			\centering
			\includegraphics[width= 0.24 \textwidth]{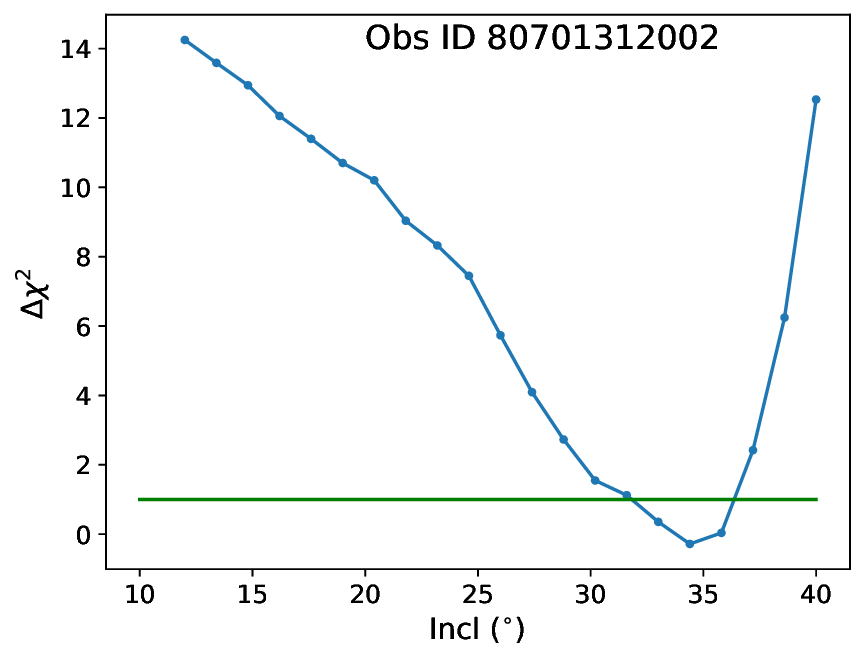}
			\includegraphics[width= 0.24 \textwidth]{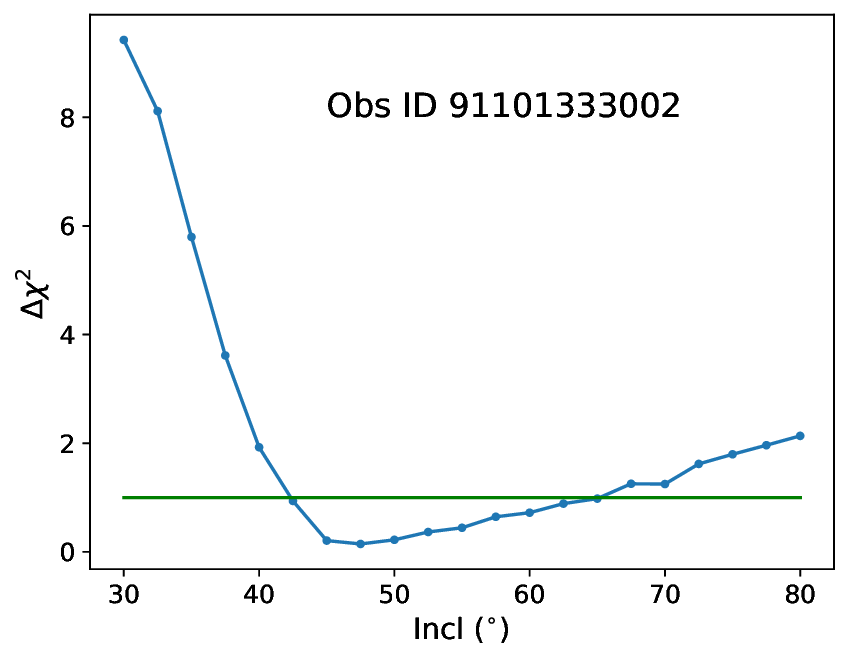}
			\includegraphics[width= 0.24 \textwidth]{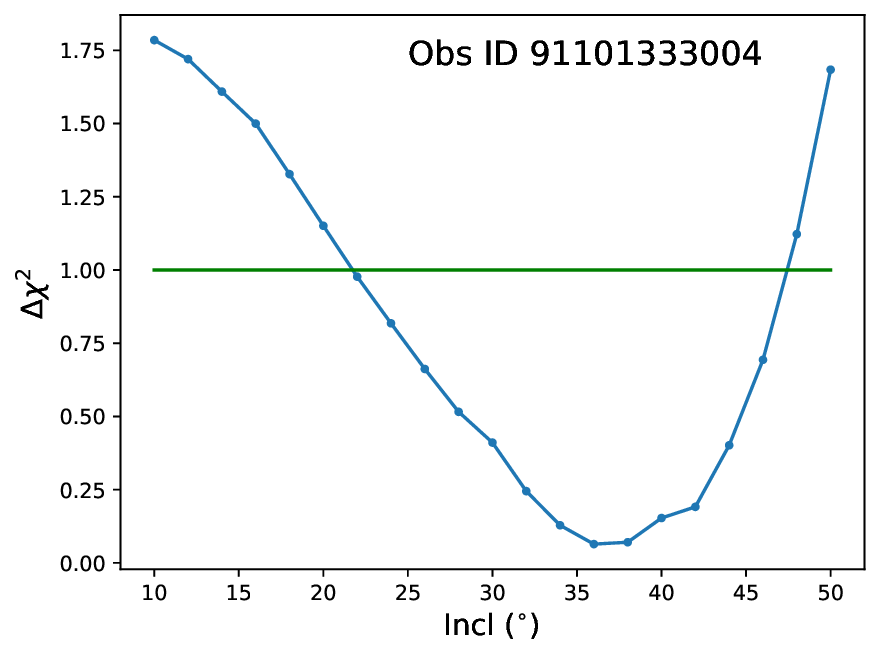}
			\includegraphics[width= 0.24 \textwidth]{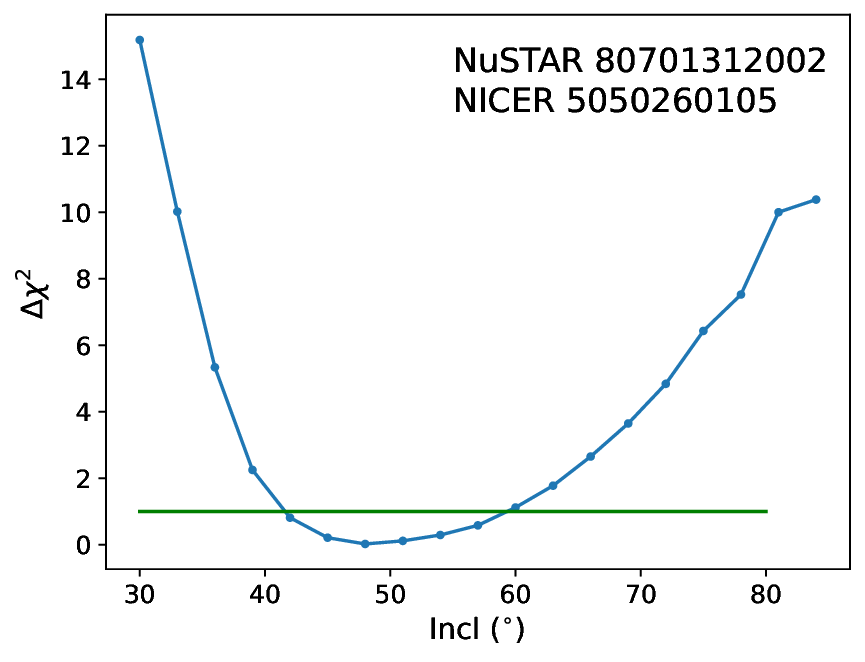}\\
			\includegraphics[width= 0.24 \textwidth]{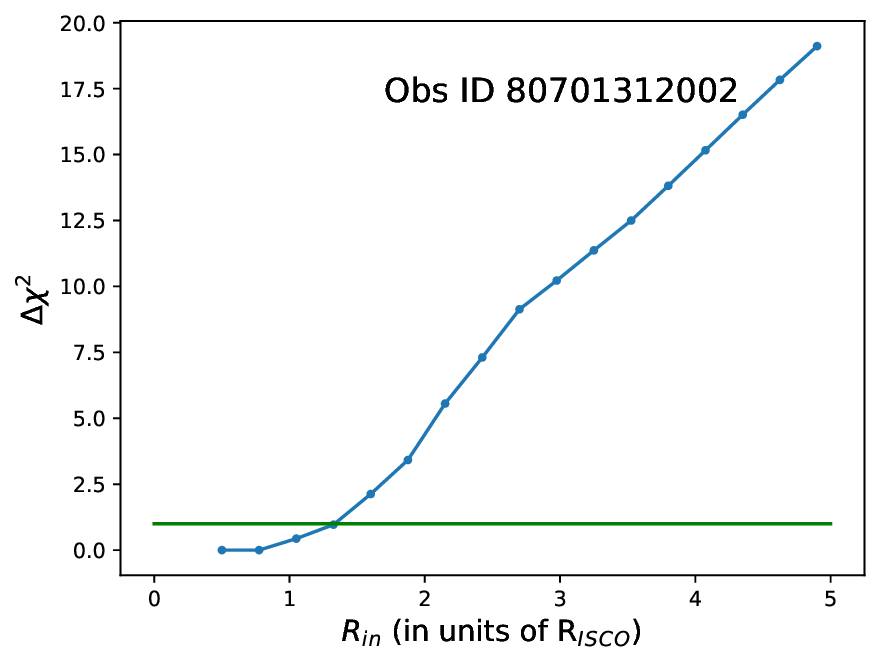}
			\includegraphics[width= 0.24 \textwidth]{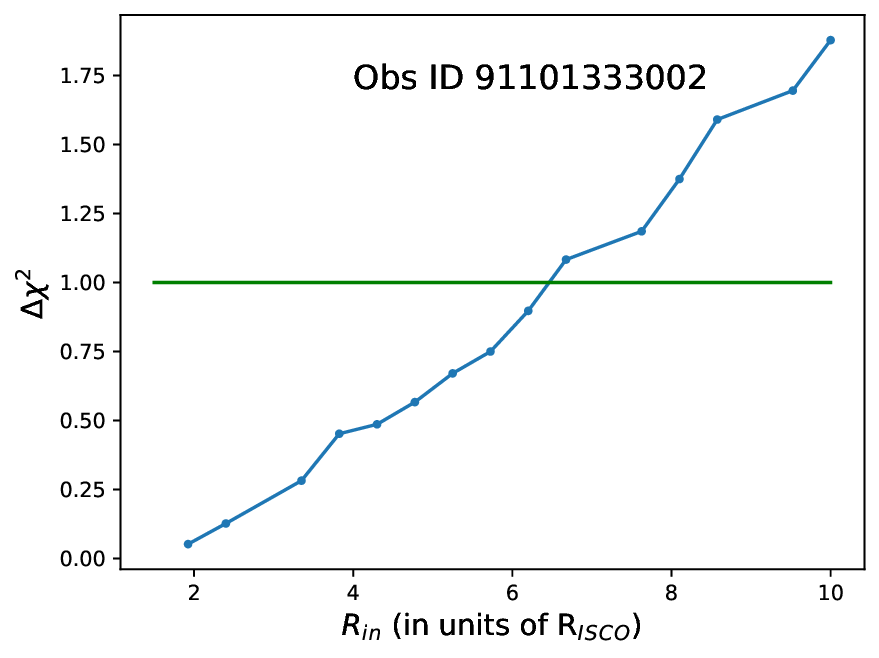}
			\includegraphics[width= 0.24 \textwidth]{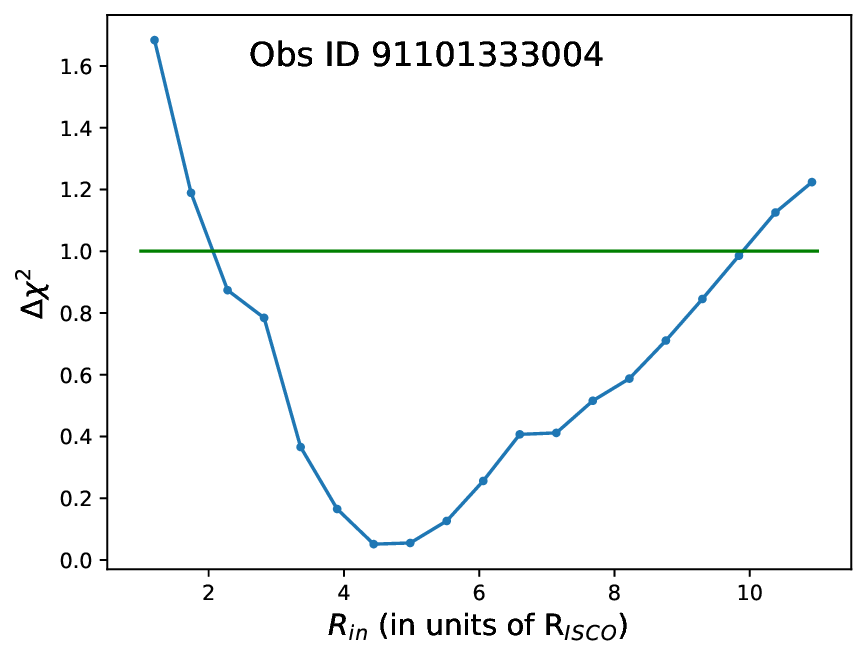}
			\includegraphics[width= 0.24 \textwidth]{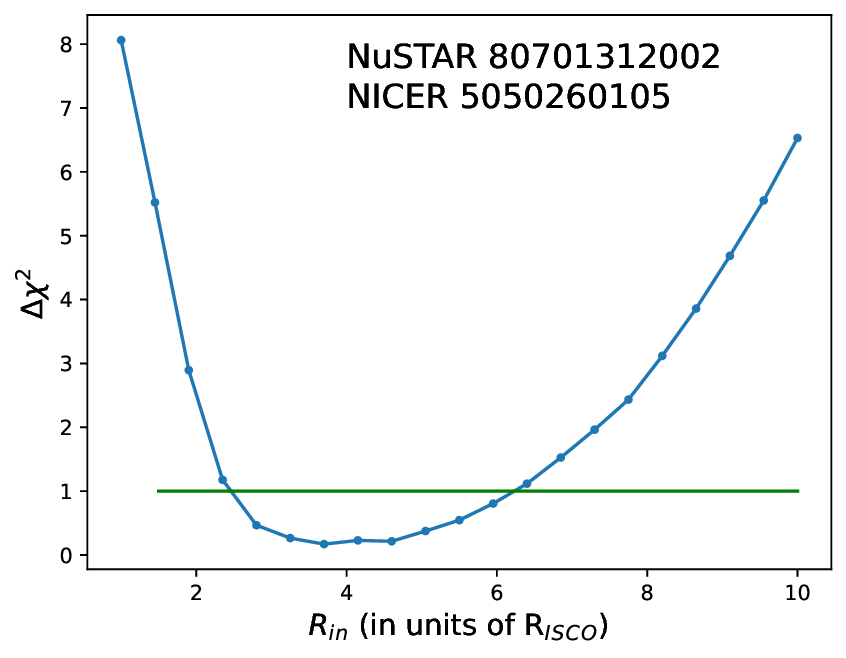}
			\caption{Variation of $\chi^2$ ($\Delta \chi^2=\chi^2 - \chi_{\rm min}^2$) as a function of model parameters is shown for the three \textit{NuSTAR} observations (Obs 1, 2 and 3) and the joint fit of \textit{NuSTAR} Obs 1 and \textit{NICER} Obs 5, labeled by their respective observation IDs. The \textit{upper panel} shows angle of inclination (\textit{Incl}) of the accretion disc, varied over 10$^{\circ}$-40$^{\circ}$, 30$^{\circ}$-80$^{\circ}$, 10$^{\circ}$-50$^{\circ}$ and 30$^{\circ}$-80$^{\circ}$ for  Obs 1, 2, 3, and the joint fit for Obs 1 and 5, respectively. The \textit{lower panel} shows the inner radius ($R_{\rm {in}}$) of the accretion disc, varied over 0-5 $R_{\rm ISCO}$, 1-10 $R_{\rm ISCO}$, 1-11 $R_{\rm ISCO}$, and 1-10 $R_{\rm ISCO}$ for Obs 1, 2, 3, and the joint fit for Obs 1 and 5, respectively.}
			\label{fig:incl_Rin}
		\end{figure*}

		The \texttt{steppar} command has been run with the three observations 1, 2, 3, and the joint fit of \textit{NuSTAR} Obs 1 and \textit{NICER} Obs 5 to test the best-fit values of the \texttt{relxillCP} model parameters, \textit{viz.}, inner radius (\textit{R$_{\rm in}$}) and angle of inclination ($Incl$) of the accretion disc, for the best-fit Model 3, \texttt{const*TBabs*(thComp*diskbb+relxillCP)}, to test the goodness of fit. The variation of $\chi^2$ ($\Delta \chi^2 = \chi^2 - \chi_{\rm min}^2$) as a function of the inclination angle and inner radius of the accretion disc is shown in the upper and lower panels of Fig. \ref{fig:incl_Rin}, respectively. These contours indicate that Model 3 provides well-constrained parameter estimates within the 1$\sigma$ confidence interval.\\
		
		\subsection{Burst spectral analysis}
		\label{sec:burst_analys}

		A burst has been observed in the 2025 September \textit{NuSTAR} observation 3. The burst took place for a duration of 50 s from 2025-09-14 11:01:46 (MJD 60932.46) from the onset of observation. As the burst evolves rapidly, a time-resolved spectral analysis is required to capture its different phases and to know the evolution of different spectral parameters during the burst.\\
		To analyse the persistent emission before the burst, we considered a time segment of 2000 s before the burst. An exposure of 2 ks was extracted from 2025-09-14 09:23:26 (MJD 60932.39) to 2025-09-14 09:56:46(MJD 60932.41). Model 2 was used for performing the spectral analysis of the pre-burst emission. The $\chi^2/dof$ was given as 83/97. Corresponding to \texttt{thComp} component, the parameters $\Gamma$ and $kT_{\rm e}$ are obtained as $1.97_{-0.04}^{+0.03}$ and $40_{-22}^{+50}$, respectively. The covering fraction was estimated to be less than 0.74. For the \texttt{diskbb} component, the parameters $T_{\rm {in}}$ and \textit{norm} were estimated to be 0.57$_{-0.08}^{+0.18}$ and $18\pm3$, respectively. We used these values as a background for the burst emission. The 0.1-100 keV flux during the persistent emission for the 2 ks segment before the burst is estimated using the distance to the source, $d\sim3.5$ kpc, as $F_{\rm pers}=3.1\times10^{-10}$ ergs cm$^{-2}$ s$^{-1}$ which corresponds to a luminosity of $L_{\rm pers}=4.5\times10^{35}$ ergs s$^{-1}$.
		
\begin{figure*}
			\includegraphics[width=0.22 \textwidth, angle=270]{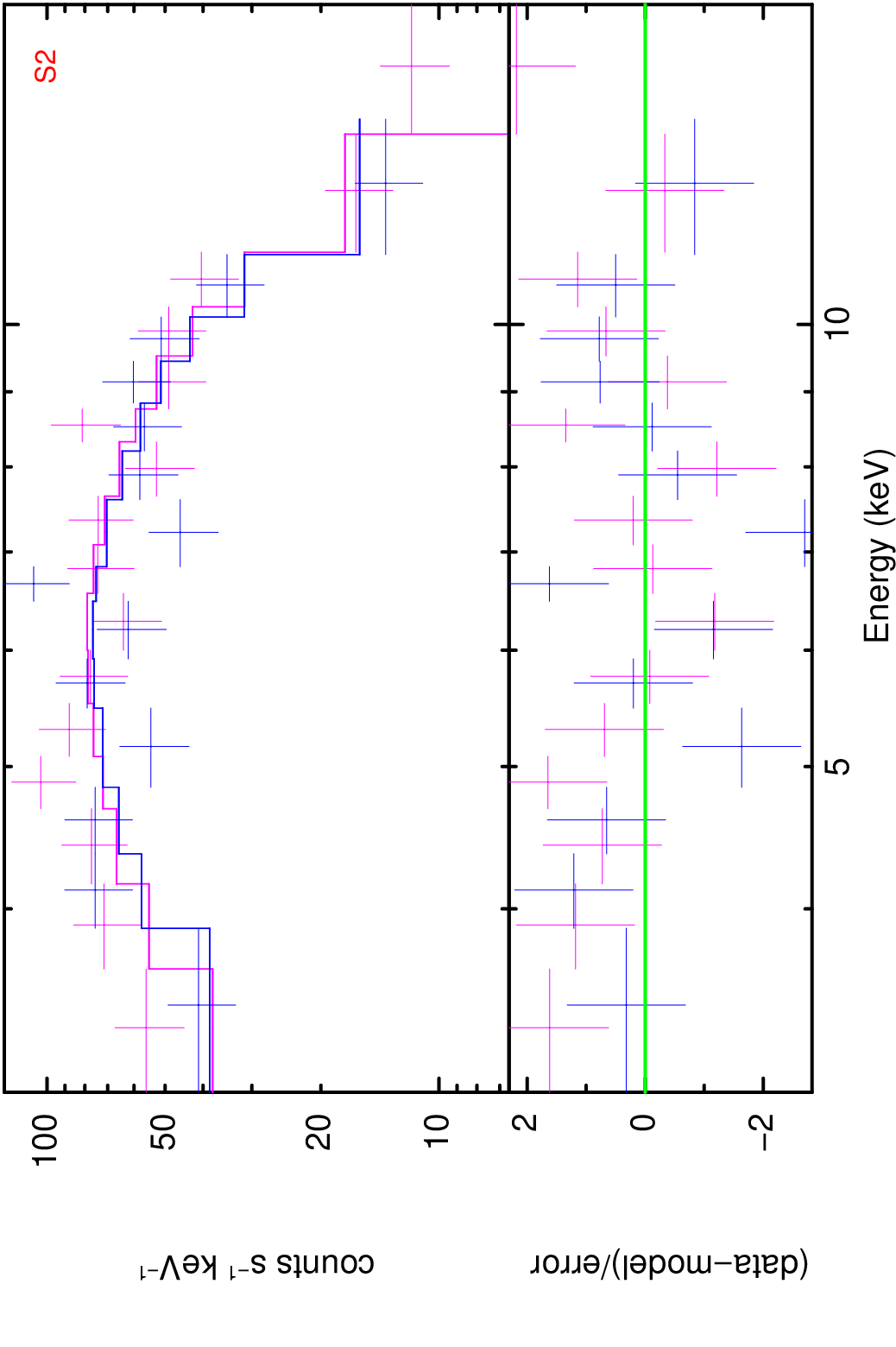}
			\includegraphics[width=0.22 \textwidth, angle=270]{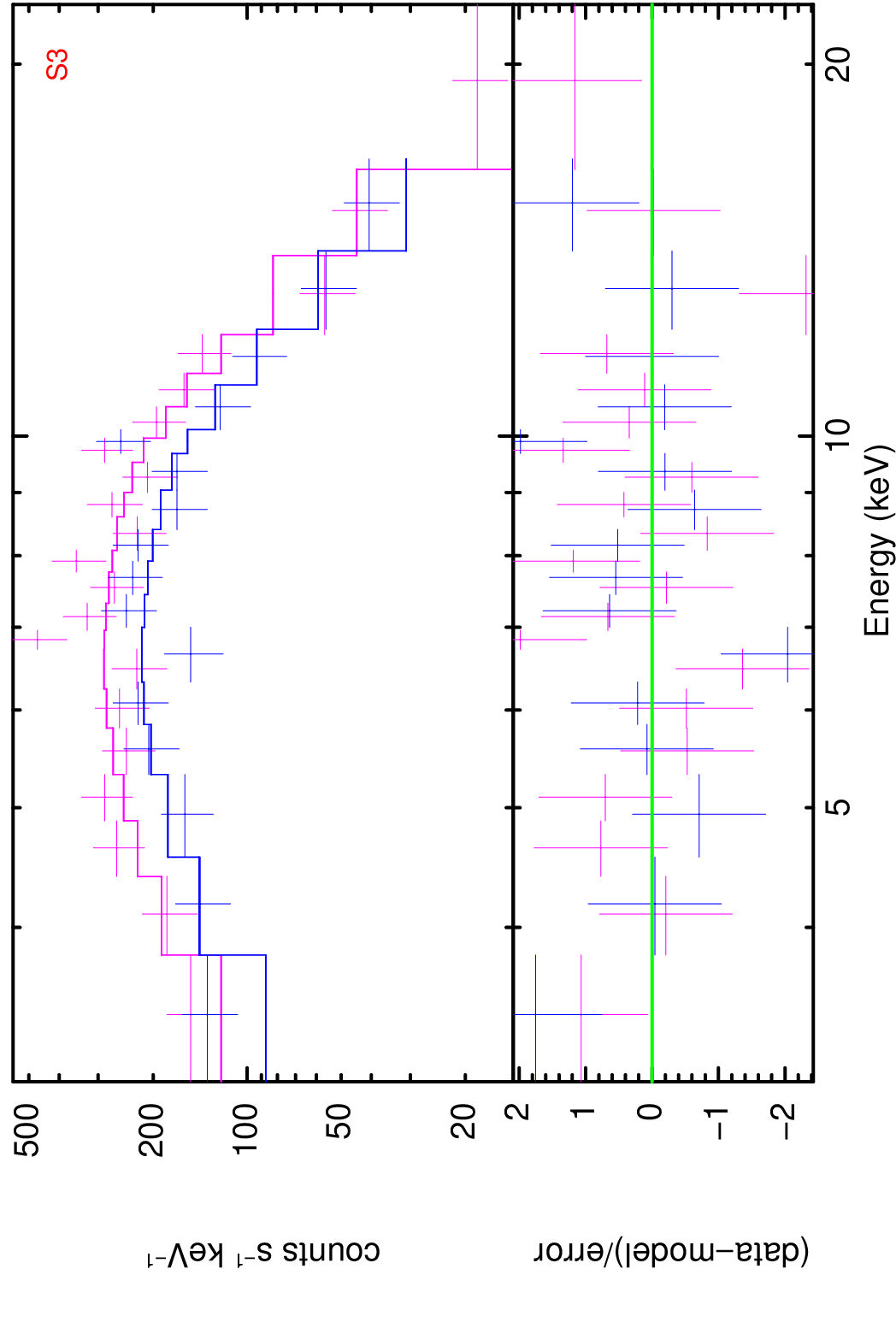}
			\includegraphics[width=0.22 \textwidth, angle=270]{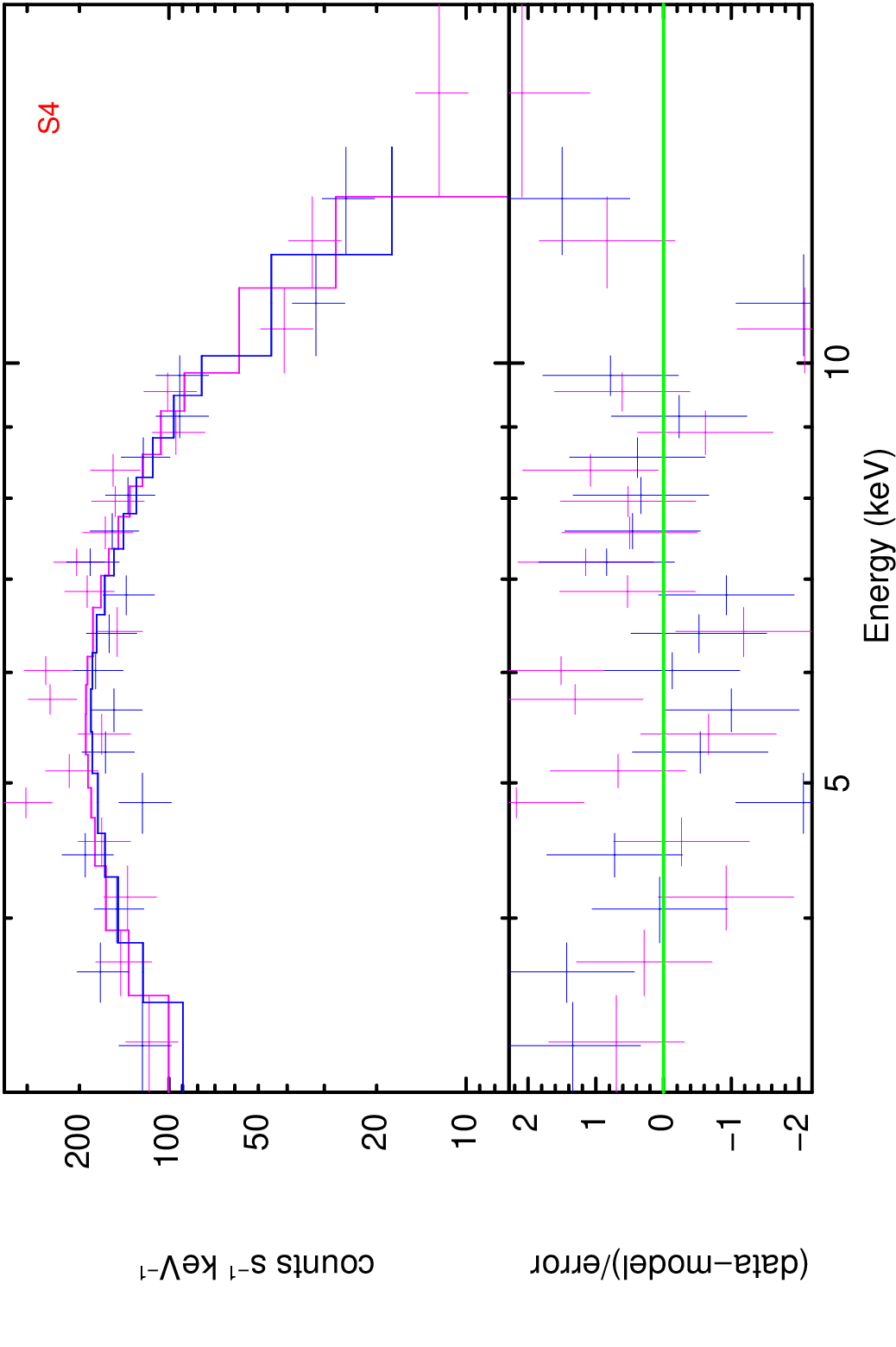}\\
			\includegraphics[width=0.21 \textwidth, angle=270]{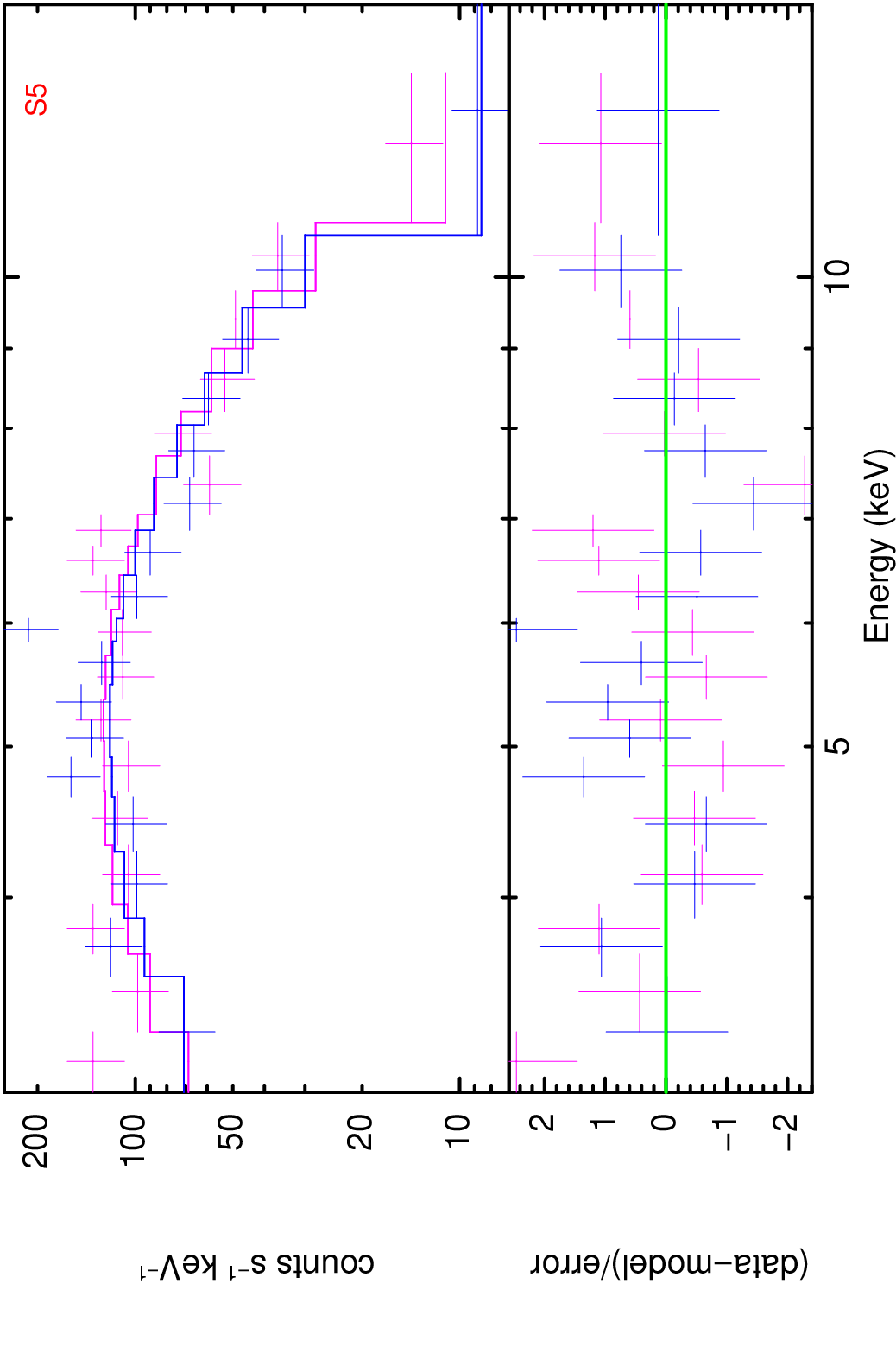}
			\includegraphics[width=0.22 \textwidth, angle=270]{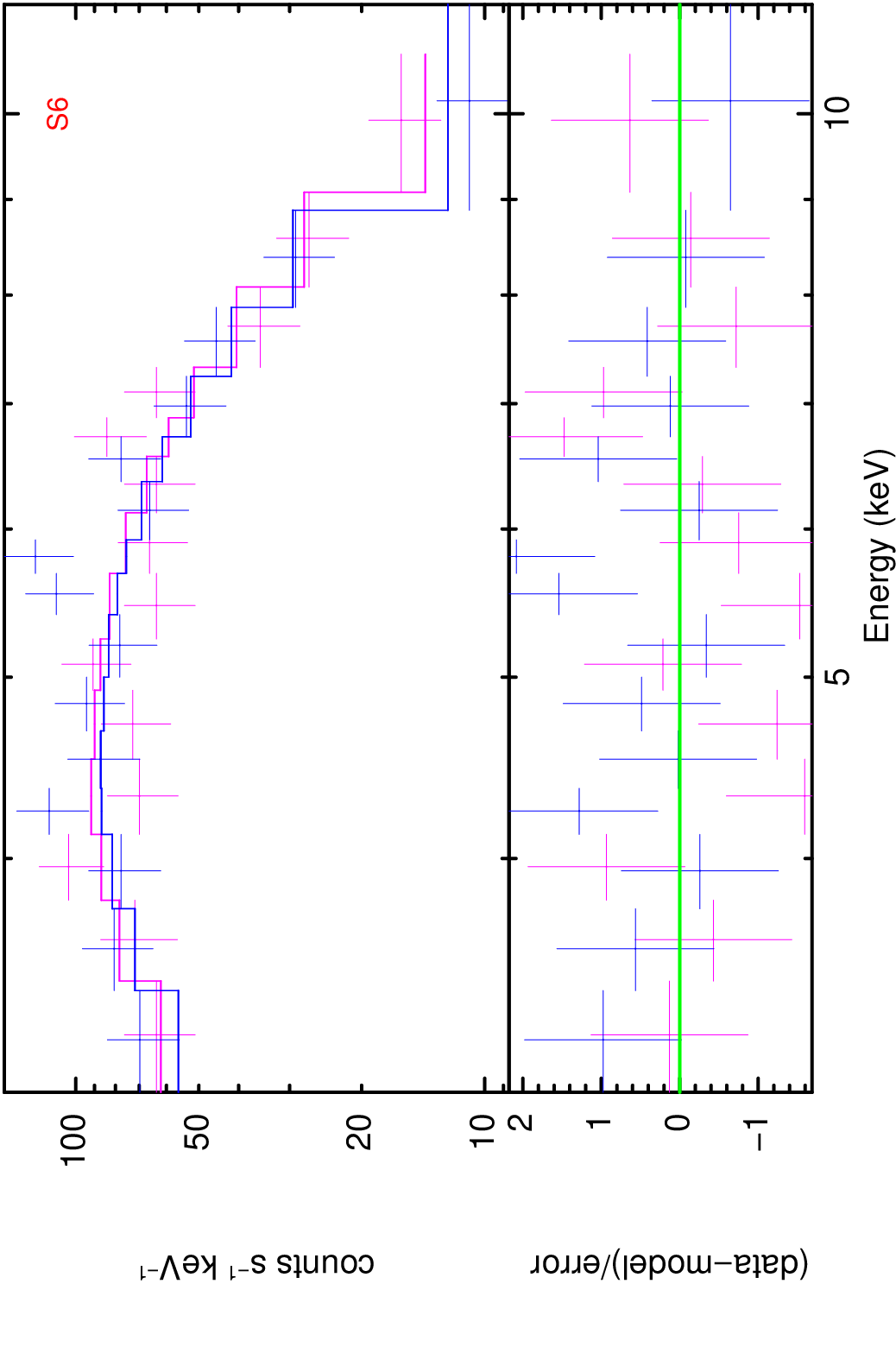}
			\includegraphics[width=0.22 \textwidth, angle=270]{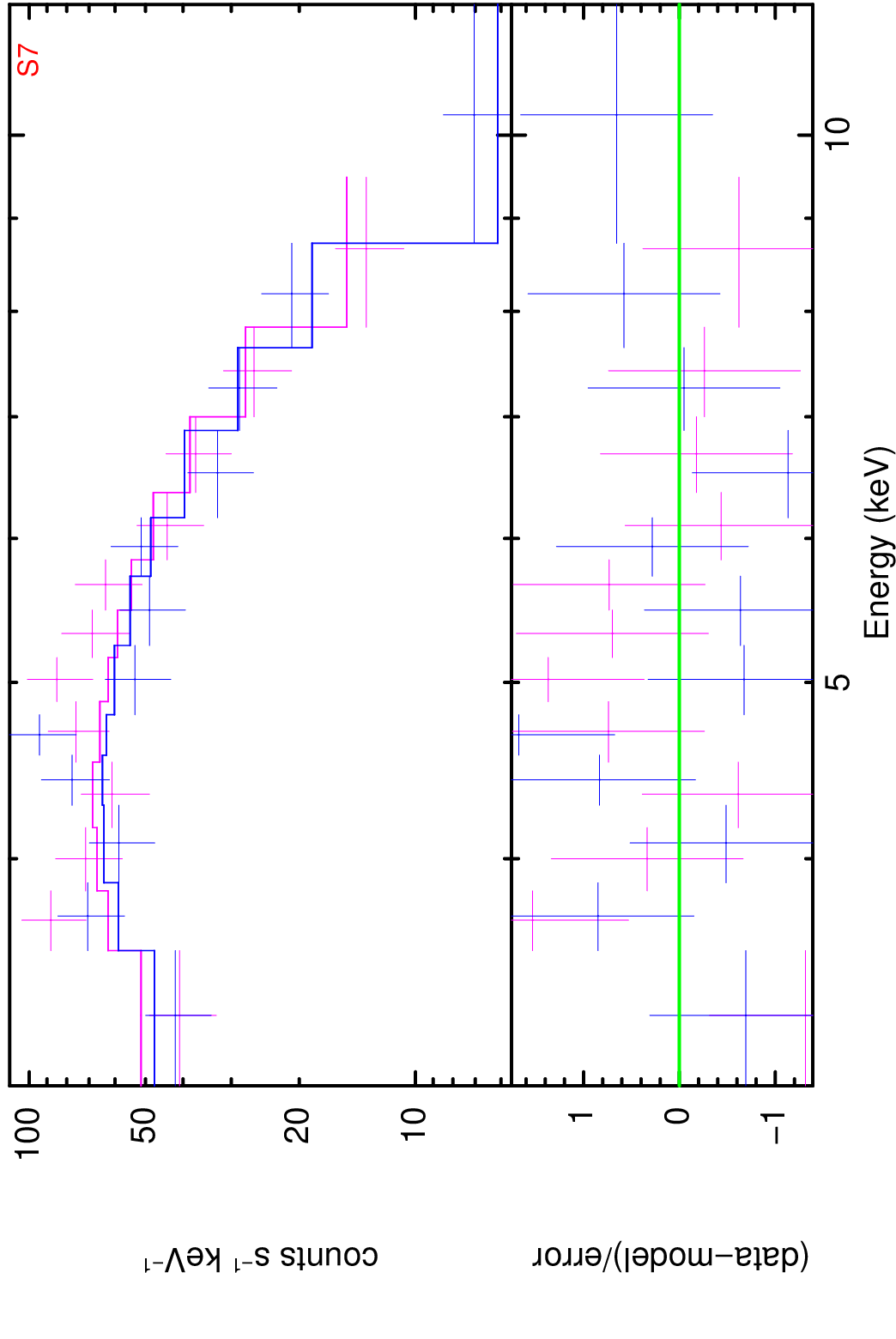}\\
			\includegraphics[width=0.21 \textwidth, angle=270]{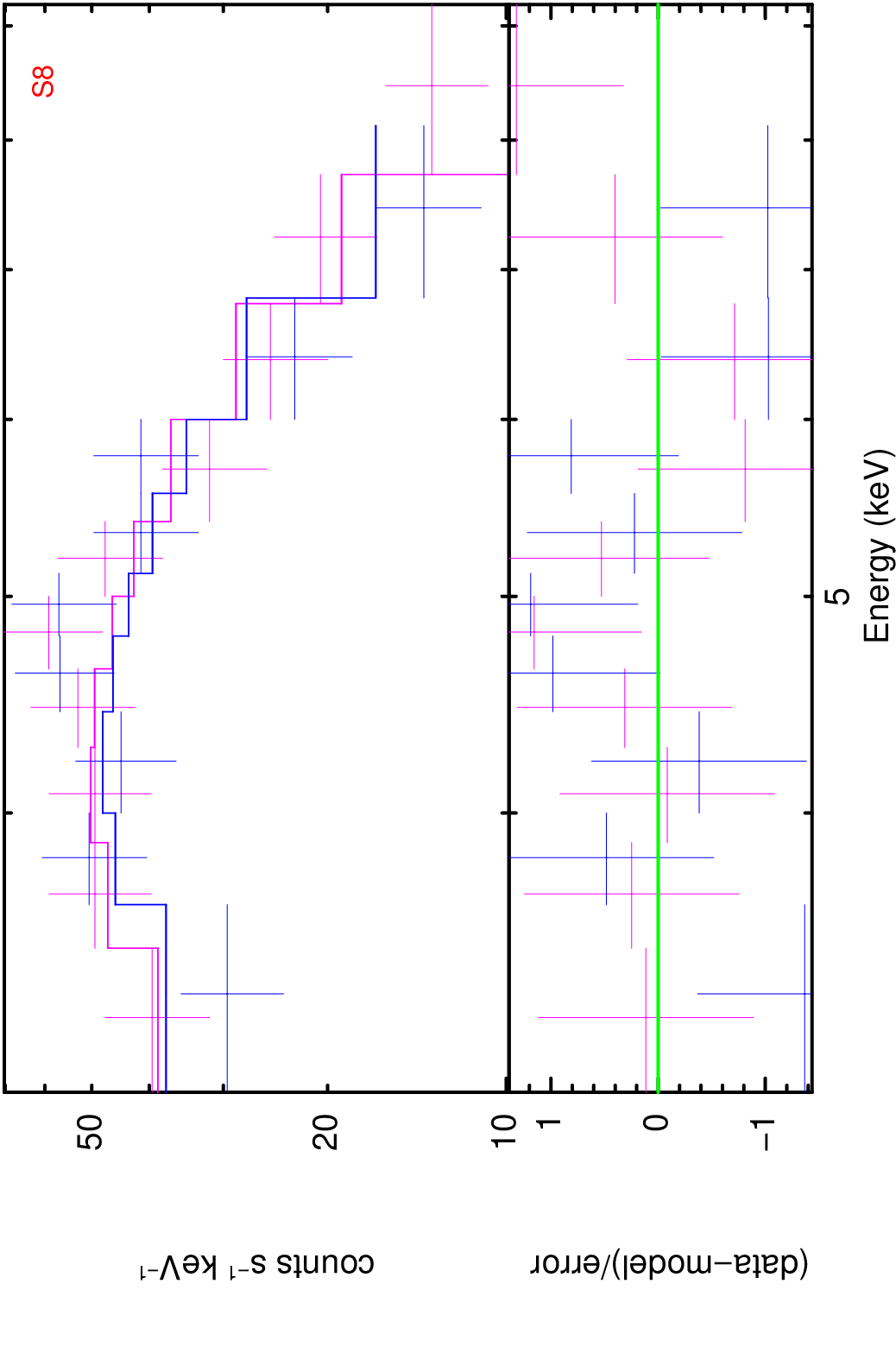}
			\includegraphics[width=0.22 \textwidth, angle=270]{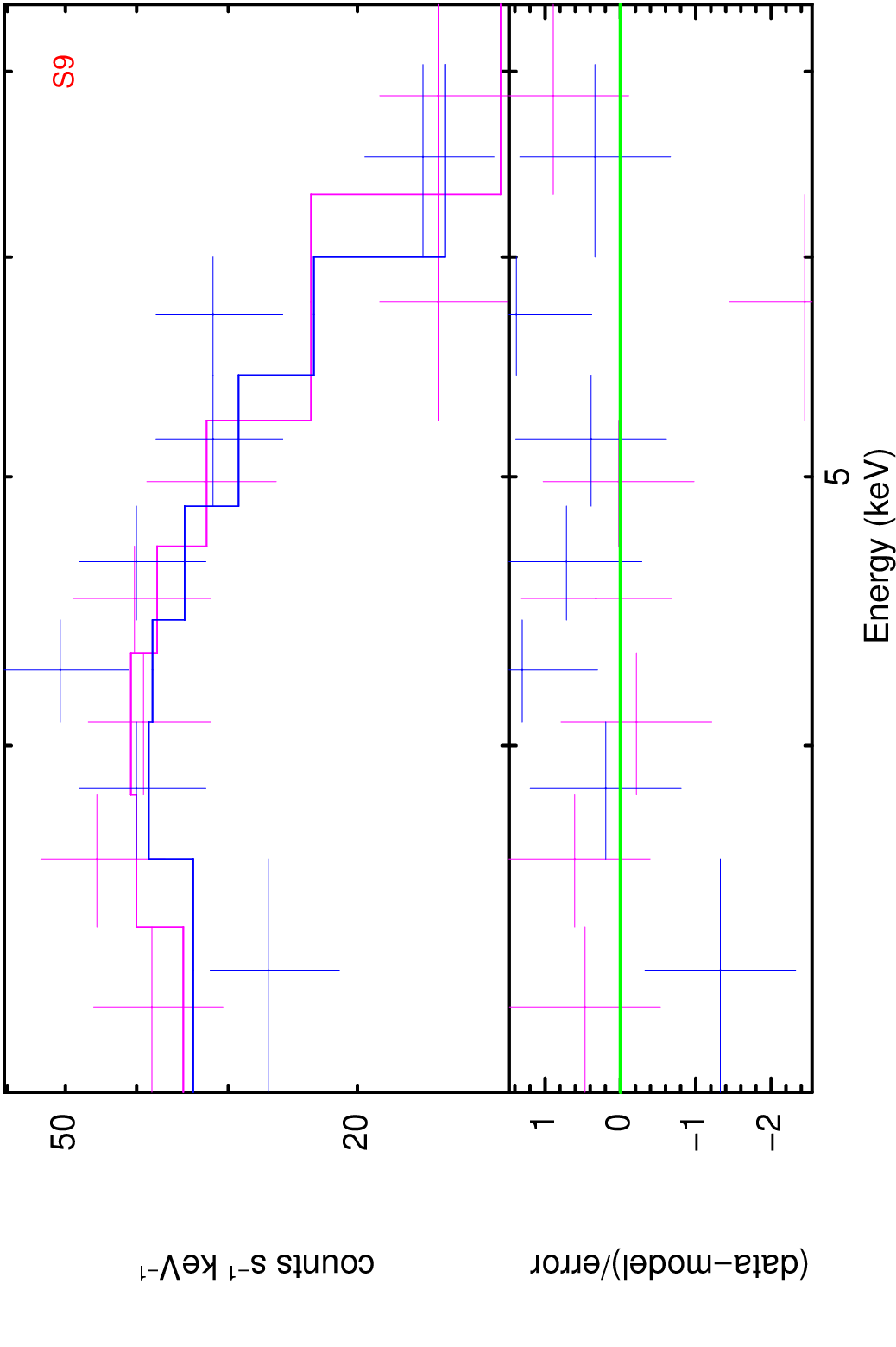}
			\includegraphics[width=0.22 \textwidth, angle=270]{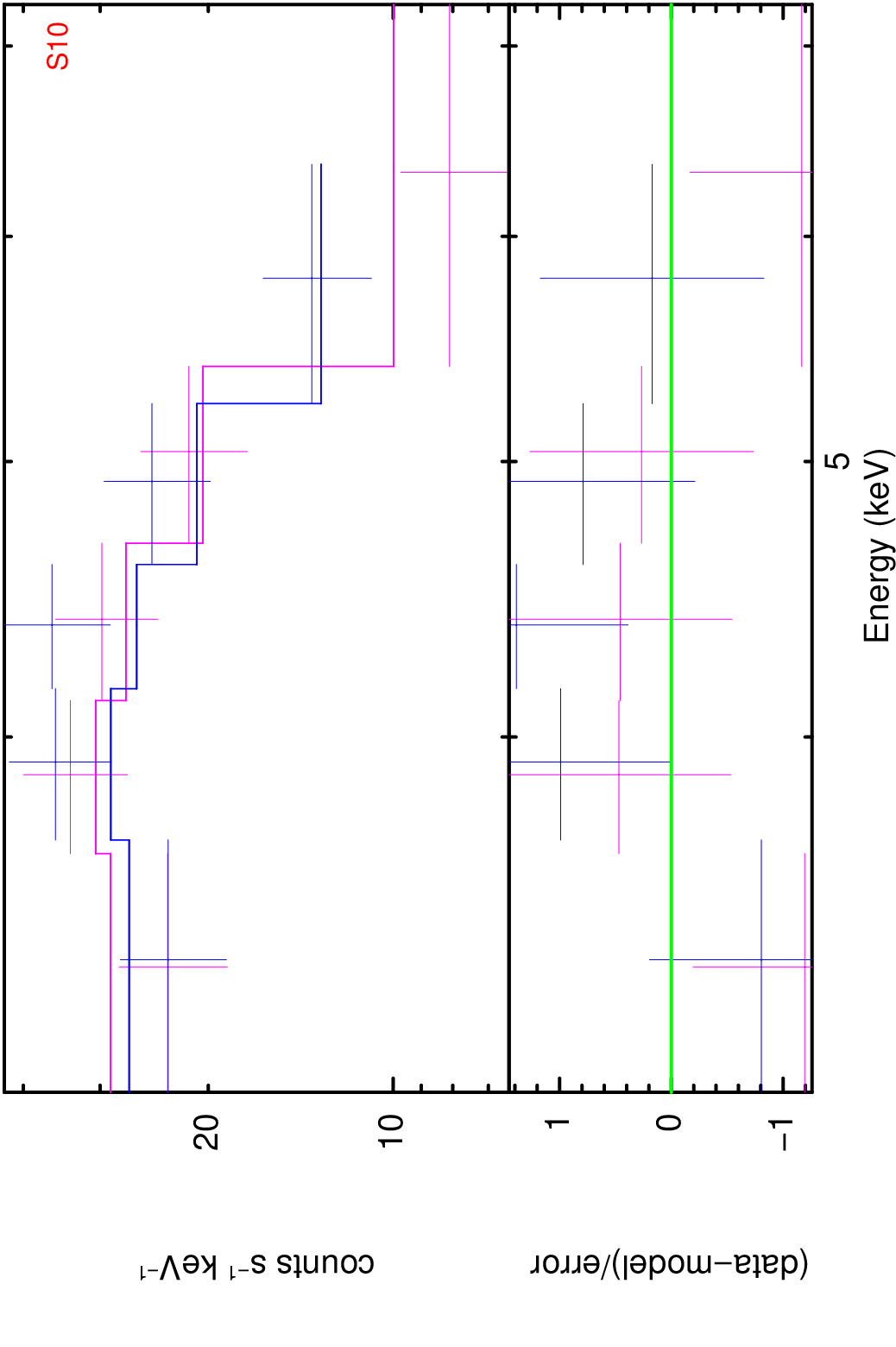}\\
			\caption{Unfolded spectral plots corresponding to the burst emission using the best fit data of model \texttt{TBabs*bbodyrad} for the \textit{NuSTAR} FPMA and FPMB corresponding to observation 3 in the energy range 3-20 keV of the source SAX J1808 for the segments S2-S10 each of 2s duration. The lower plots within each plot shows the residual plot of the particular data and model difference within 1$\sigma$ error bar.}
			\label{fig:burst_seg}
		\end{figure*}

		We have divided the burst exposure into small 2s segments, which are named as S1, S2, and so on. To ascertain the thermal emission during the burst segment, we performed spectral fit of burst segments using an absorbed blackbody model \texttt{Tbabs*bbodyrad}. The best-fit values corresponding to the segments that provided statistically acceptable fits are listed in Table \ref{tab:burst_seg}. The segments S2 and S3 correspond to the rise and peak phases of the burst, respectively. The blackbody temperature during the peak burst is $2.73\pm0.09$ keV, and the corresponding blackbody emitting radius is $5.3\pm0.3$ km. The spectral plots corresponding to the model \texttt{TBabs*bbodyrad} within 1$\sigma$ uncertainty are shown in Fig. \ref{fig:burst_seg}.\\
		The first segment (S1) did not provide sufficient statistics for reliable spectral fitting. To fetch a reliable spectral fit, we added 200s of pre-burst time segment into S1. Still, there were no adequate statistics. Hence, we have excluded S1 from analysis. Segment S1 and those after S12 did not provide adequate statistics for spectral fitting and were not included in this study. The temporal evolution of parameters like blackbody temperature, flux, and apparent radius throughout the burst exposure is shown in Figure \ref{fig:flux,kT,norm}.
		
		\begin{table*}
			
			\caption{Best-fit results for burst emission from 2025-09-14 11:01:46 to 11:02:36 (elapsed time stretch 23,136 s - 23,234 s from the beginning of observation) for the \textit{NuSTAR} FPMA and FPMB observation of the source SAX J1808 in the energy range 3-20 keV. The model for evaluating the burst emission is Model: \texttt{TBabs*bbodyrad}. All uncertainties are quoted at the 1$\sigma$ confidence level.}

			\begin{tabular}{cllcccl}
				\hline
				\parbox{3.5cm}{Time elapsed from \\2025-09-14 11:01:46\\ \centering(in (s))} & \parbox{1.2cm}{Burst\\ segment} & \parbox{1cm}{\textit{$kT_{bb}$} \\ (keV)} & Norm & Radius (km) & \parbox{1.8cm}{\textsc{flux}\\ (3-20 keV) \\  \centering{$F_{\rm b}^{\%}$}} &  $\chi^2/dof$ \\
				\hline
				1 $\pm$ 1 & S2(2s) & 2.23 $\pm$ 0.08 &  118 $\pm$ 17 & 3.79 $\pm$ 0.27 & 2.85 & 35/27 \\ 
				3 $\pm$ 1 & S3(2s) & 2.73 $\pm$ 0.09 & 229$_{-26}^{+30}$  & 5.31 $\pm$ 0.32 & 12.2 & 34/33 \\
				5 $\pm$ 1 & S4(2s) & 1.91 $\pm$ 0.05 & 436$_{-49}^{+54}$  & 7.32 $\pm$ 0.43 & 5.56 & 44/35 \\
				7 $\pm$ 1 & S5(2s) & 1.64 $\pm$ 0.05 & 457$_{-54}^{+61}$  & 7.50 $\pm$ 0.47 & 2.96 & 36/32 \\
				9 $\pm$ 1 & S6(2s) & 1.36 $\pm$ 0.04 & 588$_{-75}^{+85}$ & 8.51 $\pm$ 0.58 & 1.66  & 21/26 \\
				11 $\pm$ 1 & S7(2s) & 1.25 $\pm$ 0.04 & 637$_{-92}^{+106}$ & 8.86 $\pm$ 0.68 & 1.20 & 17/21 \\
				13 $\pm$ 1 & S8(2s) & 1.16 $\pm$ 0.05 & 617$_{-111}^{+133}$  & 8.73 $\pm$ 0.86 & 0.85 & 12/16 \\
				15 $\pm$ 1 & S9(2s) & 1.03 $\pm$ 0.04 & 723$_{-131}^{+161}$ & 9.46 $\pm$ 0.94 & 0.56 & 14/13 \\
				17 $\pm$ 1 & S10(2s) & 0.89 $\pm$ 0.05 & 1021$_{-275}^{+365}$ & 11.30 $\pm$ 1.74 & 0.36 & 6/7 \\
				19 $\pm$ 1 & S11(2s) & 0.83$_{-0.08}^{+0.05}$ & 1217$_{-297}^{+821}$ & 13.21 $\pm$ 2.59 & 0.30 & 3/5 \\
				21 $\pm$ 1 & S12(2s) & 0.81 $\pm$ 0.06 & 896$_{-306}^{+454}$ & 10.68 $\pm$ 2.18 & 0.19 & 6/4 \\
				
				\hline
				\textit{Note:} & \multicolumn{5}{l}{$^{\%}$ denotes the  unit of flux in $10^{-8}$ erg cm $^{-2}$ s $^{-1}$.}
			\end{tabular}
			\label{tab:burst_seg}
		\end{table*}
		
		\section{Discussion}
		\label{sec:disc}
		
		We have performed spectral analysis of three \textit{NuSTAR} observations of the source SAX J1808 observed on 2022 September 22, 2025 August 30, and 2025 September 14. We have included the spectral analysis of the two \textit{NICER} observations and also performed a joint spectral fit for the simultaneous 2022 August \textit{NuSTAR} and \textit{NICER} observations (Obs 1 \& 5). The average count rate for the \textit{NuSTAR} Obs 1, 2, and 3 (except bursts) in the energy range 3-79 keV is obtained as $\sim$15--18, 5--7, and 3--5 counts s$^{-1}$, respectively. The 0.1--100 keV flux, luminosity, and mass accretion rate for Obs 1, 2, and 3 are specified in Table \ref{tab:mass_acc} assuming a distance of $\sim$ 3.5 kpc \citep{galloway2006helium, 2009MNRAS.400..492I}. The 3-20 keV flux for Obs 1, 2, and 3 is $\sim$6, $\sim$2.3, and $\sim$1.4 $\times 10^{-10}$ ergs cm$^{-2}$ s$^{-1}$, respectively. The hardness--intensity diagram (shown in Fig. \ref{fig:HI_diag}) shows that the two 2025 observations occupy lower-intensity regions than the 2022 observation. The 2022 and 2025 observations are separated by an extended interval of quiescence, sampling different accretion regimes, accretion rates (as revealed by Table \ref{tab:mass_acc}), and correspond to distinct stages of different outbursts. Therefore, no temporal evolution can be inferred from the hardness-intensity diagram alone.\\
		The persistent emission was described by a thermal Comptonization model \texttt{thComp}, assuming seed photons originate from the accretion disc. Model 2, \texttt{TBabs(thComp*diskbb)}, in the energy band of 3--79 keV for Obs 1 and 3--50 keV for Obs 2 and 3, provided a good fit (Table \ref{tab:mod1,2}). The spectral features of the broad Fe emission line around 5--8 keV and the Compton hump around 15-30 keV, clearly detected, suggesting a reflection of hard photons from the accretion disc. We used Model 3 involving \texttt{relxillCP}, a reflection model, to investigate the parameters associated with the disc. The inner radius and inclination angle of the accretion disc are primary in ascertaining the time evolution of the disc. Model 3 provided these parameters within an acceptable range of uncertainty (1$\sigma$). Fig. \ref{fig:incl_Rin} shows that the best--fit values are well constrained within 1$\sigma$ uncertainty. The \textit{NICER} observations 4 and 5 were analysed using the \texttt{diskbb+powerlaw} model, which showed broad iron emission features around 6--8 keV. Furthermore, the parameters of Obs 1 and 5 were constrained through a joint spectral fit, which provided most parameter values consistent with those obtained from the individual fits.\\

\begin{figure*}
			\centering
			\includegraphics[width=0.7 \textwidth]{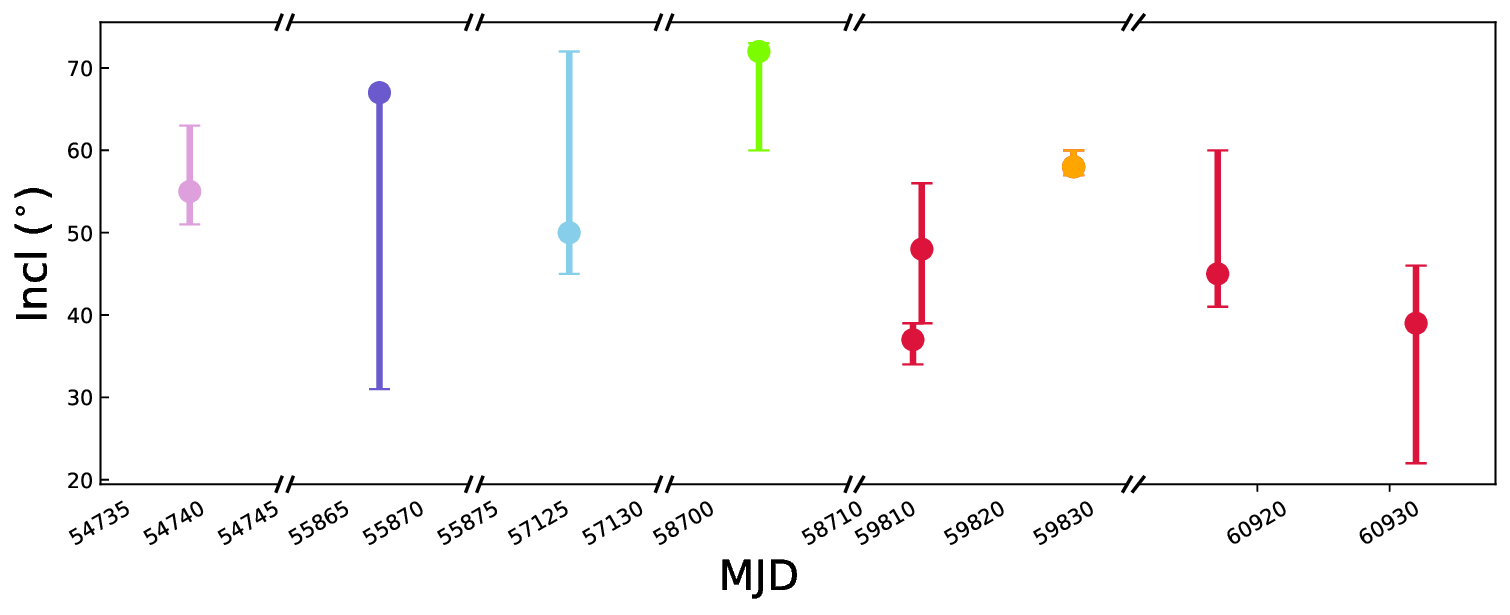}\\
			\includegraphics[width=0.7 \textwidth]{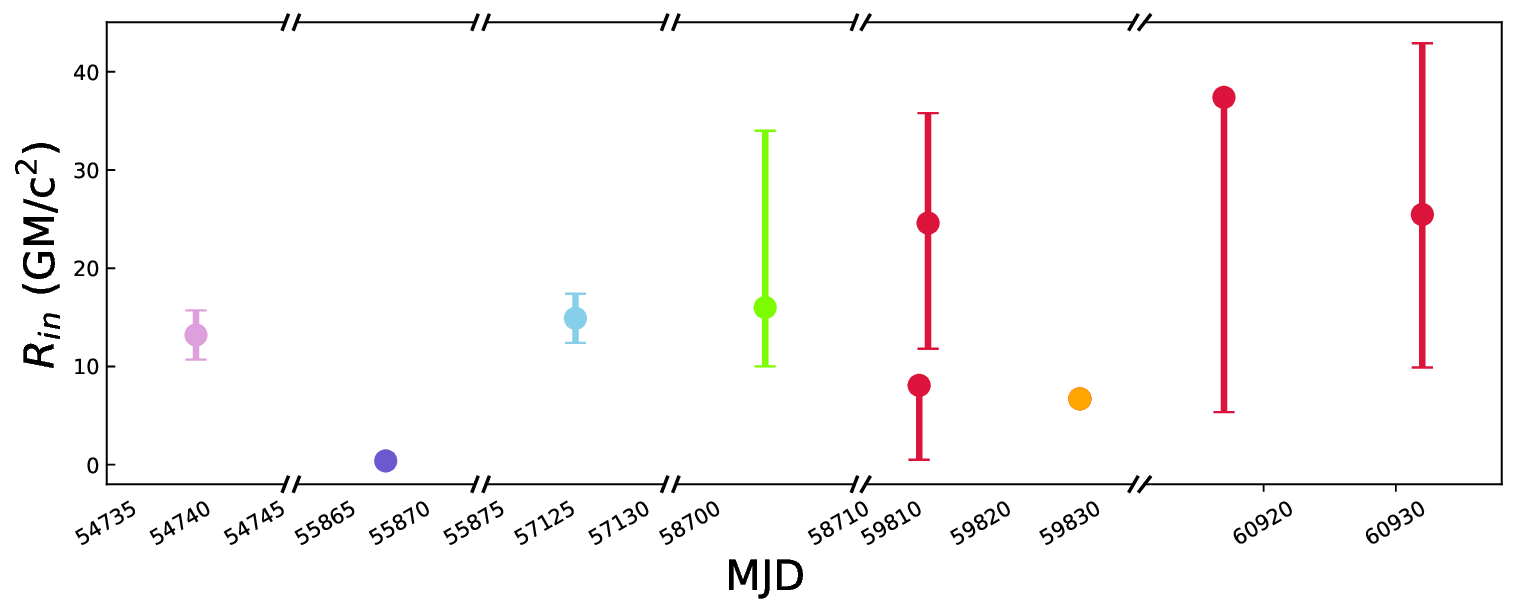}\\
			\caption{Time-evolution of the parameters - \textit{upper panel:} inclination angle, and \textit{lower panel:} inner radius of the accretion disc. The values are compiled from previous studies (listed in Table \ref{tab:prev_obs}) corresponding to the years 2008(\textit{plum}), 2011(\textit{steel blue}), 2015(\textit{sky blue}), 2019(\textit{green}) and 2022 (\textit{orange}), along with the results from this work (\textit{crimson red}).}
			\label{fig:time_evo}
		\end{figure*}

		The inner radius of the accretion disc is estimated as $\lesssim$1.5, $\lesssim$7.0, 4.8$_{-2.9}^{+3.3}$ and 4.6$_{-2.4}^{+2.1}$ in units of $R_{\rm ISCO}$, where  $R_{\rm ISCO}$ is the radius of the innermost stable circular orbit, corresponding to Obs 1, 2, 3, and the joint fit of Obs 1 \& 5. Following \citealt{1998ApJ...509..793M} we have, $R_{\rm ISCO} = \frac{6GM}{c^2} [1 - j (\frac{2}{3})^{1.5}]$, where $M$ is the mass of the neutron star and $j$ is the dimensionless angular momentum $j\simeq cJ/GM^2$ with $J$ being the angular momentum of the star. For SAX J1808, assuming the spin parameter of $j=0.2$, we obtain $R_{\rm ISCO} = 5.4 GM/c^2 \sim$ 11.1 km (assuming a NS mass of $M=1.4M_{\odot}$). Thus, the inner disc radius (within 1$\sigma$ uncertainty) is estimated to be $\lesssim$8.1, $\lesssim$37.4,  25.5$^{+17.4}_{-15.6}$ and 24.6$_{-12.8}^{+11.2}$ in units of $\frac{GM}{c^2}$ for Obs 1, 2, 3, and joint fit of Obs 1 \& 5, respectively, corresponding to inner disc radii of $\le$12, $\lesssim$55, 38$_{-23}^{+26}$, and 36$_{-19}^{+17}$ in units of $\left(\frac{M}{M_{\odot}}\right)$ km. Further, assuming M$=1.4M_{\odot}$, the inner radii are obtained as $\lesssim$17, $\lesssim$78, 53$_{-32}^{+36}$ and 51$_{-27}^{+23}$ km. We note that the inner disc radius is well constrained by the data when we consider the joint spectral fit to the simultaneous 2022 August \textit{NuSTAR} and \textit{NICER} observations. The extent of the inner radius as computed for Obs 1 with Model 3 ($\lesssim$8.1$R_{\rm g}$) is consistent with the value $\sim$6.7$R_{\rm g}$ estimated by \cite{Ballocco_2025} for the \textit{XMM-Newton} observation during the same time period. For same 2022 \textit{NuSTAR} observation, \cite{2026ApJ...999..133S} fixed the inner radius at $1 R_{\rm ISCO}\sim6 R_{\rm g}$ which agrees with our estimated result.\\
		The inclination angle of the accretion disc was estimated as $37^{\circ}{}^{+2^{\circ}}_{-3^{\circ}}$, $45^{\circ}{}^{+15^{\circ}}_{-4^{\circ}}$, $39^{\circ}{}^{+7^{\circ}}_{-17^{\circ}}$, and $48^{\circ}{}_{-9^{\circ}}^{+8^{\circ}}$ for Obs 1, 2, 3, and joint fit of Obs 1 \& 5, respectively. The inclination angle obtained by \cite{2026ApJ...999..133S} was $39.3^{\circ} {}^{+3.6^{\circ}}_{-1.9^{\circ}}$ which is comparable to the value estimated using Model 3 ($37^{\circ}{}^{+2^{\circ}}_{-3^{\circ}}$). For the \textit{XMM-Newton} observation, the inclination angle obtained was $58^{\circ} {}^{+2^{\circ}}_{-1^{\circ}}$. All these estimates lie within the range of inclination angle $36^{\circ}$-$67^{\circ}$ suggested for SAX J1808  \citep{Deloye_2008}. We note that on performing the joint fit of Obs 1 and 5 the inclination angle is also consistent in this range. The inclination angle corresponding to Obs 1, 2, and 3 is less than $67^{\circ}$ which indicates the system is viewed at moderate inclination. The variation in inclination angle across the three measurements is likely due to the model-dependent degeneracies inherent to the fitting procedure, particularly due to the cross-correlation between free parameters, rather than an intrinsic change in the orientation of the system.\\		
		The variability in the reported values of the inclination angle and the inner disc radius of the source SAX J1808 is shown in Fig. \ref{fig:time_evo}. The corresponding values reported during the 2008, 2011, 2015, 2019, and 2022 outbursts of SAX J1808, together with those estimated from the present work, are listed in Table \ref{tab:prev_obs}. The inclination angle is observed to vary approximately within $\sim30^{\circ}-60^{\circ}$, which is consistent with a moderately inclined system. However, model-dependent degeneracies may play an important role in estimating the inner disc radius in various cases.\\
		The disc ionization parameter (log $\xi$) inferred from the \texttt{relxillCP} model differs among the three \textit{NuSTAR} observations, with values of $\sim2.83$, $\lesssim1.30$, and $\sim3.22$ for Obs 1, Obs 2, and Obs 3, respectively. The joint fit of Obs 1 \& 5 yields $\log\xi\sim2.45$, which is consistent with the value obtained for Obs 1. The ionization parameter is significantly lower in Obs 2, whereas the values obtained for Obs 1 and Obs 3 are consistent within their uncertainties. Additionally, in the \texttt{relxillCP} model, the component $\log N$ refers to the electron density of the disc in cm$^{-3}$. As this parameter is not well constrained by the data, we performed the spectral analysis by fixing $\log N$ to an appropriate value commonly adopted for neutron star low-mass X-ray binaries. Following the earlier study of \cite{2026ApJ...999..133S}, we fixed $\log N=$18 cm$^{-3}$ for all three observations. Additionally, by performing a joint fit of \textit{NuSTAR} Obs 1 and \textit{NICER} Obs 5, we obtained $\log N\sim18.5$. A similar trend was observed for the iron abundance parameter ($A_{\rm Fe}$), which was not well constrained during the individual spectral analyses but was estimated as $A_{\rm Fe}\sim1$ during the joint fit. As the iron abundance was not constrained during the individual spectral analyses of the three \textit{NuSTAR} observations, we have fixed the iron abundance parameter to 1 in those fits. The upper bound of the electron temperature (k$T_{\rm e}$) was not well constrained during the joint spectral fit.\\

		\begin{figure}
			\centering
			\includegraphics[width=0.48\textwidth]{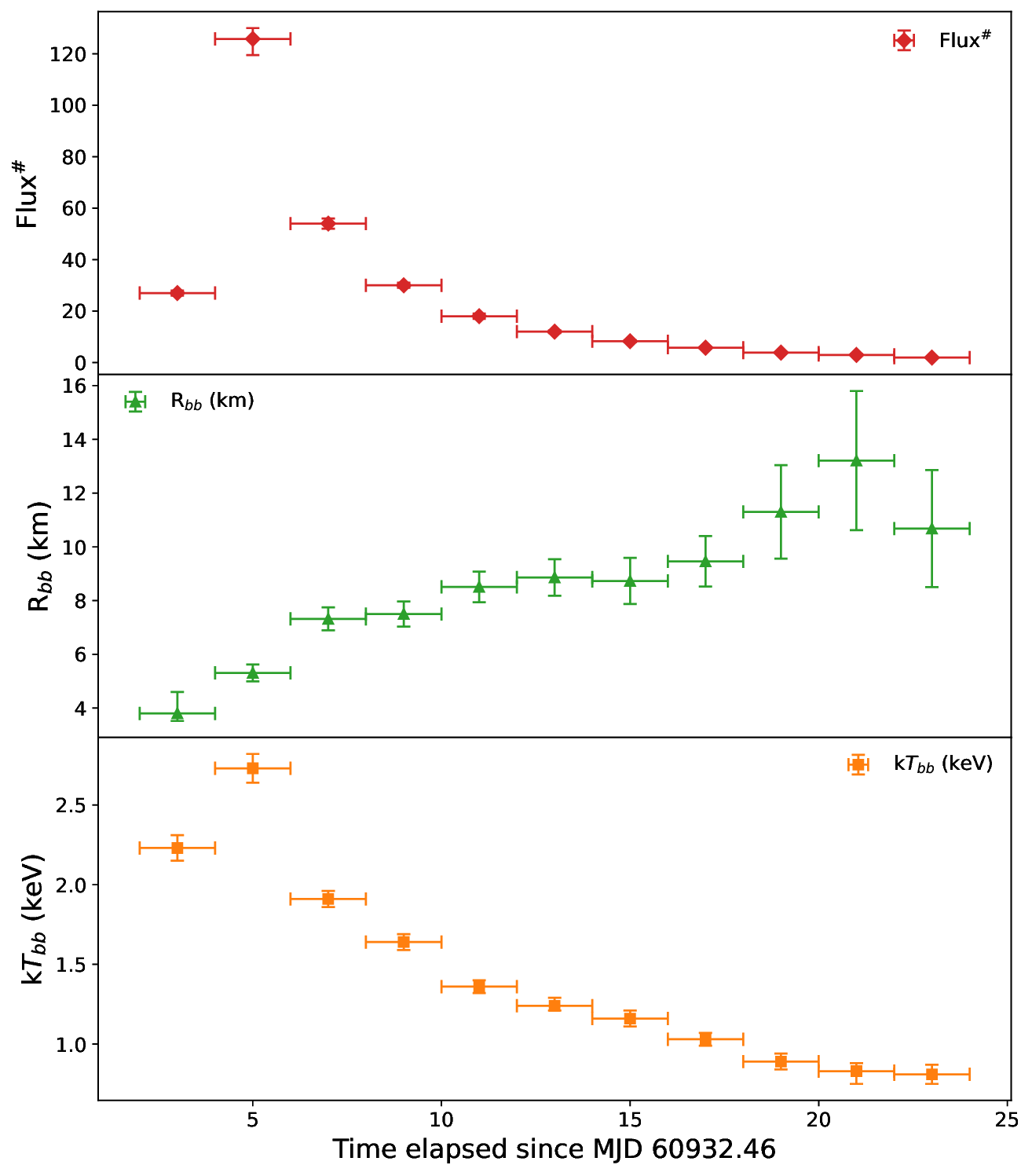}
			\caption{Evolution of the spectral parameters for blackbody emission during the burst (Table \ref{tab:burst_seg}) for the \textit{NuSTAR} FPMA and FPMB observation of the source SAX J1808 during 2025 September. \text{$^{\#}$} Unit of flux in 10$^{-8}$ erg cm$^{-2}$ s$^{-1}$.}
			\label{fig:flux,kT,norm}
		\end{figure}
		
		\begin{table*}
			\caption{Luminosity and mass accretion rate corresponding to the extended 0.1-100 keV bolometric flux obtained using the \texttt{cflux} component in \textsc{XSPEC}, for the \textit{NuSTAR} observations 1, 2, 3, and the joint fit of \textit{NuSTAR} Obs 1 and \textit{NICER} Obs 5.}
			\begin{tabular}{c c c c}
				\hline
				Observation Year and Month & \parbox{2.4cm}{Flux$^\%$ ($F$) (0.1-100 keV)} & \parbox{2.5 cm}{Luminosity$^{*}$ \\ \centering $ L = 4 \pi d^2 F$ } & \parbox{4cm}{Mass accretion rate \\ \centering $\dot{M} = \frac{LR}{GM}$ \\ \centering(g s$^{-1}$)} \\
				\hline
				2022 Aug (80701312002) & 12.7  & 18.6  & 9.9 $\times 10^{15}$ \\
				2025 Aug (91101333002) & 4.5 & 6.7 & 3.6 $\times 10^{15}$ \\
				2025 Sep (91101333004) & 2.9 & 4.2  & 2.3 $\times 10^{15}$ \\
				2025 Sep (91101333004 burst) & 45.4 & 66.6 & 3.6 $\times 10^{16}$ \\
				2022 Aug (Joint fit of Obs 1 $\&$ 5) & 13.5 & 19.8 & 10.6 $\times 10^{15}$\\
				\hline
				\textit{Note:} & \multicolumn{3}{l}{$^\%$ denotes the unit of flux in 10$^{-10}$ ergs cm$^{-2}$ s$^{-1}$} \\
				& \multicolumn{3}{l}{$^{*}$ denotes the unit of luminosity in 10$^{35}$ ergs s$^{-1}$} \\
			\end{tabular}
			\label{tab:mass_acc}
		\end{table*}
		
		The electron temperatures were estimated as 83$_{-35}^{+3}$, 85$_{-37}^{+10}$, and 100$_{-51}^{+10}$ keV for Obs 1, 2, and 3, respectively. The electron temperatures inferred for the two 2025 observations are consistent within their uncertainties. Both observations yield higher best-fit electron temperatures than the 2022 observation, although the values are associated with large uncertainties. Within these uncertainties, the results are consistent with the presence of a hot Comptonizing corona during the 2025 observations. The photon indices obtained using Model 1 are $\sim$$1.93$, $\sim$$1.99$, and $\sim$$2.03$ for Obs 1, 2, and 3, respectively. The spectral analysis reveals a dominant Comptonized component, indicating hard spectral characteristics. In the 2022 observation, the source has been reported to exhibit similar hard spectral characteristics \citep{2026ApJ...999..133S}. Furthermore, the 2025 observations occupy lower-intensity regions of the hardness-intensity diagram and correspond to lower inferred luminosities and mass-accretion rates than the 2022 observation. Despite these differences, the broadband spectral analysis indicates that the source remained in a hard spectral state across all three observations. Thus, the three observations probe different luminosity and inferred mass-accretion-rate regimes associated with distinct outbursts, rather than successive phases of a single evolutionary sequence.\\
		From \cite{2009MNRAS.400..492I}, the mass accretion rate is calculated using $\dot{M}=\frac{LR}{GM}$, where L is the luminosity noted in Table \ref{tab:mass_acc}, and we have assumed the mass and radius of the source as $M=1.4 M_\odot$ and $R=$10 km. During the 2022 August observation, the mass accretion rate is obtained as $\sim$9.9$\times10^{15}$ g/s. For 2025 August and September observations, the mass accretion rate is obtained as $\sim$3.6$\times10^{15}$ g/s and $\sim$2.3$\times10^{15}$ g/s, respectively. From these estimated mass accretion rate values, no temporal evolution can be inferred, as the observations were obtained during different outbursts and are separated by an interval of quiescence. Instead, we note that they sample different accretion regimes and different stages of distinct outbursts. The upper limit obtained for the inner disc radius for the 2022 observation indicates that the inner accretion disc extends close to the innermost stable circular orbit, consistent with efficient accretion as shown by the high mass accretion rate. In contrast, the source is inferred to be in a lower accretion regime during 2025 observations.\\
		
		\begin{table*}
			\caption{Best-fit values of $R_{\rm in}$ and $Incl$ based on analysis of X-ray observations of SAX J1808 during its outbursts.}
			\begin{tabular}{l c c c c c}
				\hline
				\parbox{2 cm}{Observation year and month} & MJD  & Incl ($^{\circ}$) & R$_{\rm in} (GM/c^2)$ & \parbox{2.9cm}{X-ray Observatory} & Reference \\
				\hline
				&&&&&\\
				2008 October 02 & 54741 & $55^{+8}_{-4}$ & $13.2\pm2.5$ & \parbox{2.9cm}{\textit{Suzaku} and \textit{XMM-Newton}} & \cite{Cackett_2009}   \\
				2011 November 04 & 55869 & $\lesssim$ 67 & $0.38\pm0.24$ & \textit{Chandra} & \cite{Patruno_2013} \\
				2015 April 15 & 57127 & $50^{+22}_{-5}$ & $14.9\pm2.5$  & \textit{NuSTAR} & \cite{2019MNRAS.483..767D} \\
				2019 August 10 & 58705 & $72^{+1}_{-12}$ & $16^{+18}_{-6}$ & \textit{NuSTAR} &  \cite{2026ApJ...996...73B} \\
				2022 August 22 & 59813 & $37^{+2}_{-3}$ & $\lesssim$ 8.07 & \textit{NuSTAR} & This work \\
				2022 August 23 & 59814 & 48$_{-9}^{+8}$ & 24.6$_{-12.8}^{+11.2}$ & \parbox{2.9cm}{\textit{NuSTAR} $+$ \textit{NICER}} & This work \\
				2022 September 09 & 59831 & $58^{+2}_{-1}$ & $\lesssim$ 6.7& \textit{XMM-Newton} & \cite{Ballocco_2025} \\
				2025 August 30 & 60917 & $45^{+15}_{-4}$ & $\lesssim$ 37.4 & \textit{NuSTAR} & This work \\
				2025 September 14 & 60932 & $39_{-17}^{+7}$ & 25.5$^{+17.4}_{-15.6}$ & \textit{NuSTAR} & This work \\
				\hline
			\end{tabular}
			\label{tab:prev_obs}
		\end{table*}
		
		The inferred inner disc radii for Obs 2 (upper limit) and Obs 3 indicate truncation of the accretion disc. Disc truncation at larger radii has been observed for several NS LMXBs \citep{king20162016, Fabian_2017}. There may be a different reason for this. Firstly, the disc truncation is associated with the morphological and physical evolution of the inner accretion flow occurring during a spectral state transition \citep{1997ApJ...489..865E}. In our case, such a reason for disc truncation is highly unlikely, as no prominent spectral change was observed during the 2025 \textit{NuSTAR} observations (\textit{i.e.}, Obs 2 and Obs 3).
		
		Secondly, magnetic pressure exerted on the disc is considered responsible for a cause of disc truncation \citep{Sunyaev_1975}. Following \citealt{2009MNRAS.400..492I, 2010ApJ...720..205C}, the magnetic moment can be calculated using 
		\begin{equation}
			\begin{split}
				\nonumber
				\mu = 3.5 \times 10^{23} x^{7/4} k_{\rm A} ^{-7/4} \left( \frac{M}{M_{\odot}}\right)^2 \left( \frac{f_{\rm ang}}{\eta} \right)^{1/2} \\  \left( \frac{F}{10^{-9} {\rm erg} \, {\rm cm}^{-2} \, {\rm s}^{-1}}\right)^{1/2} \left( \frac{D}{3.5 {\rm kpc}}\right)
			\end{split}
		\end{equation}
		
		where $\eta$ is the accretion efficiency, $f_{\rm ang}$ is the anisotropy correction factor, $F$ is the bolometric flux (see Table \ref{tab:mass_acc}) and $k_{\rm A}$ is the geometric coefficient. Following \citealt{2010ApJ...720..205C}, we have assumed $\eta=0.1$, $f_{\rm ang}=1$ and $k_{\rm A}=1$. The factor $x$ is inferred from $R_{\rm in}$ = $\frac{x GM}{c^2}$. For the three observations, we estimate the parameter $x$ by taking the upper limits on the inner disc radius for Obs 1 and Obs 2, and best-fit value for Obs 3 and the joint fit of Obs 1 and 5. Then, the upper limits on $x$ are $\sim$8 and $\sim$37 for Obs 1 and Obs 2, respectively, while for Obs 3 and the joint fit of Obs 1 and 5, we obtain $x$ as $26_{-16}^{+17}$ and $25_{-13}^{+11}$, respectively. Assuming a distance to the source as $d=$3.5 kpc, the magnetic moment obtained corresponding to the upper limit of Obs 1 and 2 are $\sim$9$\times10^{25}$ G cm$^3$, $\sim$82$\times10^{25}$ G cm$^3$, respectively, and for Obs 3 and the joint fit of Obs 1 and 5, we obtain $34_{-27}^{+50}\times$10$^{25}$ G cm$^3$ and $66_{-48}^{+62}\times$10$^{25}$ G cm$^3$, respectively. Assuming $R=10$ km, the corresponding magnetic field can be estimated using $B=\frac{\mu}{R^3}$ as $\sim$9$\times$10$^7$ G for Obs 1, and $\sim$8$\times10^8$ G for Obs 2, 3$_{-2}^{+5}$$\times$10$^8$ G for Obs 3, and $7_{-5}^{+6}\times$10$^8$ G for the joint fit of Obs 1 and 5. The estimated magnetic fields lie above the range of minimum magnetic field ($B_{\rm min}\sim2\times10^7$ G) for an AMXP and are consistent with the previously reported value of $\sim 7\times10^7$ G for SAX J1808 \citep{2018MNRAS.480..692P}. As the time span of observations is just three years, it is highly unlikely that any significant intrinsic change in the neutron star magnetic field or the geometry of the system has occurred. The differences in the magnetic field are primarily a consequence of the uncertainties in the inferred inner disc radius, which arise from parameter degeneracies and the limited statistical constraints of the reflection spectral modeling.
		
		Thirdly, the decrease in the mass accretion rate may be a possible explanation for the disc truncation \citep{Popham_2001}. As the mass accretion rate decreases, the pressure of the infalling matter decreases. The accretion disc drifts away from the Keplerian value as the magnetic pressure dominates at larger radii \citep{1988MNRAS.231..325W}. Consequently, as the mass-accretion rate decreases, the magnetosphere expands, which can push the inner edge of the accretion disc to larger radii and truncate the disc.\\	
		To assess the extension of the magnetosphere, the magnetospheric radius is compared with a critical value, the co-rotation radius \citep{ghosh1979accretiona}. The co-rotation radius is the radial distance where the neutron star's spin frequency equals the Keplerian angular frequency of the accreting plasma. The co-rotation radius of the source can be computed using $R_{\rm co}=(\frac{GM}{\Omega^2})^{1/3}$ where $\Omega=2\pi f$ is the angular velocity corresponding to the pulsation frequency ($f$). For $f$=401 Hz and $M=1.4M_{\odot}$, we get $R_{\rm co}\sim$31 km. The magnetospheric radius is given as $R_{\rm m} \simeq \left( \frac{\mu^4}{2G M \dot{M}^2}\right)^{1/7}$ where $\dot{M}$ is the mass accretion rate. Substituting the corresponding values of magnetic moment and mass accretion rate, for Obs 1, 2, and 3, we obtained the value of the magnetospheric radius to be $\sim$30 km, $\sim$138 km, and 94$_{-58}^{+65}$ km, respectively. During Obs 1, the upper limit of the magnetospheric radius is less than, although very near, the co-rotation radius ($R_{\rm m}<R_{\rm co}$), which indicates that the magnetosphere lies inside and close to the co-rotation radius. The accretion disc allows matter to fall inwards onto the neutron star. The source is in the accretion regime. For Obs 2, the accretion rate decreases, and the upper limit of the magnetospheric radius exceeds the co-rotation radius. As we have obtained the upper limit only for this observation, any concluding remark on the accretion state for Obs 2 is unsubstantiated. For Obs 3, we see that the lower bound on the magnetospheric radius ($\sim$37 km) exceeds the co-rotation radius ($\sim$31 km), implying $R_{\rm m}>R_{\rm co}$. However, if we estimate the inner disc radius at 90\% confidence level ($\Delta\chi^2=2.7$), it returned a value of $4.8_{-3.6}^{+12.5}$$R_{\rm ISCO}$ $(25.5_{-19.2}^{+67.0} R_{\rm g})$ for Obs 3. The inferred value of the lower limit of $R_{\rm in}$ corresponds to $R_{\rm m}\sim$23 km, which is smaller than the co-rotation radius. This indicates that the result $R_{\rm m}>R_{\rm co}$ is valid only at the 1$\sigma$ confidence level. If the magnetosphere exceeds the co-rotation radius, the matter cannot accrete onto the neutron star. Additionally, the inferred mass accretion rate for Obs 3 is lower than that for Obs 2. Therefore, at the 1$\sigma$ confidence level, the combination of a lower accretion rate along with the magnetospheric radius possibly suggests that the source may have approached the propeller regime.\\
		
		During the 2025 September observation, a burst of $\sim$50 s duration was observed (Fig. \ref{fig:burst_seg}). The rise time, peak time, and decay time (defined earlier in Sec. \ref{sec:lc}) of the burst are 1 s, 3 s, and 35 s, respectively. The total integrated time of the burst, obtained as a summation of these three components, is $t_{\rm int}$ = 39 s. The burst fluence ($f_{\rm b}$), defined as the total energy radiated during the burst per unit area (in erg cm$^{-2}$), is calculated from the bolometric flux over the burst duration ($E_{\rm b}$) \citep{1993SSRv...62..223L}. From Table \ref{tab:mass_acc}, using the 0.1-100 keV bolometric flux ($E_{\rm b}$=4.5 $\times$ $10^{-9}$ ergs cm$^{-2}$ s$^{-1}$) for the burst, the burst fluence is estimated as $f_{\rm b} = E_{\rm b} \times t_{\rm int} =$ 1.8 $\times$ 10$^{-7}$ ergs cm$^{-2}$ \citep{galloway2008thermonuclear}.
		
		The ratio between the persistent emission to the burst emission ($\alpha$) is given as $\alpha=\frac{c_{\rm bol} F_{\rm pers} {\Delta t_{\rm rec}}}{f_{\rm b}}$, where $c_{\rm bol}$ is the bolometric correction factor for the source SAX J1808 \citep{galloway2008thermonuclear}, $F_{\rm pers}$ is the bolometric flux in the energy region 0.1-100 keV for the persistent spectra of the 2000 s segment taken before the burst, $f_{\rm b}$ is the burst fluence and $\Delta t_{\rm rec}$ is the recurrence time for the burst. For the 2025 September observation of the burst, $F_{\rm pers}$ and $f_{\rm b}$ as 3.1$\times10^{-10}$ ergs cm$^{-2}$ s$^{-1}$ and 1.8$\times10^{-7}$ ergs cm$^{-2}$, respectively. Since there is only one burst observed, we assume the lower limit on the recurrence time as the time elapsed before the burst since the beginning of observation, $\Delta t_{\rm rec}\gtrsim$ 23,134 s. Then, we obtain a lower limit of $\alpha\gtrsim 84.5$. The $\alpha$ value within 40 to 100 signifies helium-rich burning \citep{Cumming_2004}. Furthermore, the burst fuel composition can be inferred from $Q_{\rm nuc}$, which is the nuclear energy released per nucleon for material with solar abundances (\citealt{galloway2008thermonuclear}). For a neutron star with $M=1.4 M_{\odot}$ and radius $R=10$ km, $Q_{\rm nuc}$ (in units of 4.4 MeV nucleon$^{-1}$) is given by $Q_{\rm nuc} = \frac{44}{\alpha}$. Then, we obtain $Q_{\rm nuc}\sim0.5$, suggesting helium burning, consistent with \cite{galloway2006helium}.\\
		The source distance was estimated using $d=\sqrt{\frac{R_{\rm NS}^2 \sigma T^4}{f_{\rm x}}}$, where $R_{\rm NS}$ is the estimated radius of the neutron star, $T$ is the estimated temperature during the burst, and $f_{\rm x}$ is the flux during the burst \citep{DiSalvo2023aoc}. For the peak burst segment (S3), from Table \ref{tab:burst_seg} we can estimate $T$ as $\sim$3.2$\times$10$^7$ K and the blackbody emitting radius was obtained from the \texttt{bbodyrad} norm, $norm = \frac{R_{NS}^2}{D_{10}^2}$, as 5.3$\pm$0.3 km. The bolometric flux during S3 is $\sim$1.2$\times$10$^{-7}$ ergs cm$^{-2}$ s$^{-1}$. Thus, the distance to the source SAX J1808 is estimated as $3.7\pm0.4$ kpc, consistent with the measurement of \citealt{galloway2006helium}.\\
		Following \cite{galloway2008thermonuclear}, the ignition depth ($y_{\rm ign}$, in units of 10$^8$g cm$^{-2}$) for a neutron star of radius $R_{\rm NS}$ and gravitational redshift $z=0.31$ can be calculated using 
		\begin{equation}
			\begin{split}
				\nonumber
				y_{\rm ign} &= \frac{E_{\rm burst}(1+z)}{4 \pi R_{\rm NS}^2 Q_{\rm nuc}} \\
				\nonumber
				&= 3.0 \times 10^8 \left(\frac{f_{\rm b}}{10^{-6} \, {\rm ergs} \, {\rm cm}^{-2}}\right) \left(\frac{d}{10 \, {\rm kpc}}\right)^2  \left(\frac{1+z}{1.31}\right) \\
				\nonumber
				&\left(\frac{R_{\rm NS}}{10 \, {\rm km}}\right)^{-2}\times\left(\frac{Q_{\rm nuc}}{4.4 \, {\rm MeV} \, {\rm nucleon}^{-1}}\right)^{-1}
			\end{split}
		\end{equation}
		and is obtained as $\sim$0.14$\times10^8$ g cm$^{-2}$. The low ignition depth indicates that the burst is triggered close to the NS surface.\\		
		From Figure \ref{fig:flux,kT,norm}, we observe that the blackbody flux increases rapidly to a maximum during the peak of the burst and then decreases continuously following the decay phase of the burst. The blackbody temperature follows a similar trend, increasing during the rise and decreasing throughout the cooling tail. The blackbody emitting radius continually increases gradually up to the late decay phase and then decreases. The late-time decrease in the emitting radius (after segment S11) indicates a contraction of the blackbody emitting area due to the cooling of the neutron star surface. The Eddington luminosity for the source having $M \sim 1.4 M_{\odot}$ is $L_{\rm Edd} = 1.8\times10^{38}$ ergs s$^{-1}$. During the burst peak segment, using the 3-20 keV bolometric flux in Table \ref{tab:burst_seg} and the distance to the source as $d = 3.7\pm0.4$ kpc, we obtain $L_{\rm burst} \sim (2.0\pm0.1)\times10^{38}$ ergs s$^{-1}$, which exceeds the Eddington luminosity. As the luminosity estimated from the peak flux is compatible with the Eddington luminosity, the observed burst could be considered an Eddington-peaking burst. However, time-resolved spectroscopy revealed that the changes in temperature and radius during the burst did not exhibit PRE-like behavior.\\
		
		The detection of a type-I burst confirms that accreted matter reaches the neutron star surface despite the disc truncation. \cite{king20162016} has reported a similar instance in Aql X-1, where evidence of accretion is observed through a Type-I X-ray burst, but X-ray pulsations are absent. Truncation of the accretion disc does not imply annihilation of the inflow of matter into the neutron star, as accreting material may be channeled along field lines to the neutron star polar caps \citep{2005ApJ...623.1051D}. To be consistent with the lack of pulsations, the emission region must be larger, or it must be aligned with the spin axis \citep{2009ApJ...706..417L}.
		
		The comparison between magnetospheric and co-rotation radii suggests that the source may have approached the propeller regime by 2025 September, within 1$\sigma$ uncertainty. Although the inflow of matter from the disc may be partially inhibited, the accreted matter can still reach the polar region of the neutron star by magnetic channeling. A helium-powered Type-I X-ray burst occurs when a sufficient column depth of helium accumulates under the conditions of high temperature and pressure. At lower mass accretion rates, steady burning of hydrogen via the CNO cycle leaves a helium layer on the NS surface, which subsequently ignites to helium-rich bursts. However, if the system transitions into a strong propeller regime, the accretion onto the neutron star will be halted, preventing further accumulation of fuel and thereby stopping burst activity. The decline in mass accretion rate during 2025 observations is therefore consistent with the source moving towards a state of quiescence. Furthermore, we need future observations with sensitive X-ray observatories during the low-luminosity phase to verify whether the source enters a sustained propeller regime and investigate the corresponding evolution of its burst activity.
		
		\section{Data Availability}
		We used \textit{NuSTAR} and \textit{NICER} archival data from the \texttt{NASA} \texttt{HEASARC} database, accessed through the \textsc{XAMIN Search} portal (https://heasarc.gsfc.nasa.gov/xamin/).
		
		\section{Acknowledgements}
		This research has made use of the \textit{NuSTAR} data analysis software \texttt{NuSTARDAS}, jointly developed by the ASI Space Science Data Center (SSDC, Italy) and the California Institute of Technology (Caltech, USA). Additionally, \texttt{NICERDAS} was utilized for \textit{NICER} data reduction and analysis. For the analysis of archival data, we deeply acknowledge \texttt{HEASOFT} and \texttt{CALDB}. This research has made use of \textit{MAXI} data provided by \textit{RIKEN}, \textit{JAXA}, and the \textit{MAXI} team. We acknowledge the use of public data from the \textit{Swift/BAT} transient monitor/survey. The research work is supported by the non-NET fellowship grant of Visva-Bharati University. ASM would like to thank the Inter-University Centre for Astronomy and Astrophysics (IUCAA) for the facilities extended to him under their Visiting Associate Programme. We acknowledge the reviewer whose valuable comments have resulted in an overall improvement of the revised manuscript.
		
		
		\bibliography{SAX_apj1}{}
		\bibliographystyle{aasjournalv7}
		
		
		
	\end{document}